\documentclass[preprint,authoryear,times,12pt]{elsarticle}

\usepackage[latin1]{inputenc}
\usepackage[german,american]{babel}
\usepackage{babelbib}

\usepackage[T1]{fontenc}
\usepackage{times}
\usepackage{helvet}
\usepackage{latexsym} 

\usepackage{amsmath} 
\usepackage{amssymb} 
\usepackage{amsxtra} 
\usepackage{bm} 
\usepackage{mathtools}

\usepackage[OMLmathsfit]{isomath}

\def\msbi#1{\mathsfbfit{#1}}

\usepackage{enumitem}

\usepackage[all]{xy}

\usepackage{graphicx} 
\usepackage{color} 
\usepackage{pdfpages}
\usepackage[percent]{overpic}
\usepackage{pgf,tikz}

\definecolor{darkred}{rgb}{0.5,0.0,0.0}
\definecolor{darkgreen}{rgb}{0.0,0.5,0.0}
\definecolor{darkblue}{rgb}{0.0,0.0,0.5}

\graphicspath{{Figures/}}

\usepackage{float}

\usepackage[font=small,labelfont=rm]{caption}
\usepackage{subcaption}

\DeclareCaptionLabelFormat{mylabel}{#1 #2.\hspace{1mm}}
\usepackage{epic,eepic}

\journal{...}

\begin{document}

\begin{frontmatter}

\title{
An approach to encode divergence-free stress fields 
in neural approximations based on stress potentials
} 

\author{Mohammad S. Khorrami$^{1}$, 
Pawan Goyal$^{2}$,
Soroush Motahari$^{1}$, 
David Oexle$^{2}$, 
Jaber R. Mianroodi$^{1}$, 
Bob Svendsen$^{1,3}$, Peter Benner$^{2}$, Dierk Raabe$^{1}$}
\address{${}^{1}$Circular Metallurgy and Alloy Design,\\
Max-Planck Institute for Sustainable Materials, 
D\"usseldorf, Germany
\\[2mm]
${}^{2}$Computational Methods in Systems and Control Theory,\\ 
Max-Planck Institute for Dynamics of Complex Technical Systems, 
Magdeburg, Germany
\\[2mm]
${}^{3}$Material Mechanics, RWTH Aachen University, Aachen, Germany}

\begin{abstract}

The purpose of the current work is the development of an approach 
to account for quasi-static mechanical equilibrium in empirical (i.e., data-based) 
models for the stress field employing neural approximations (NAs), which 
include neural networks (NNs) and neural operators (NOs), in particular 
Fourier NOs (FNOs). 
Rather than including such constraints from physics in the loss 
function as done in the (now standard) physics-informed approach, 
the current approach incorporates or "encodes" such constraints 
directly into the architecture of the NA. 
As a result, both NA training and output are physically constrained in the 
physics-encoded approach, in contrast to the physics-informed approach, 
in which only training is physically constrained. 
For the current constraint of divergence-free stress, a novel encoding 
approach based on a stress potential is proposed. 

As a "proof-of-concept" example application of the current approach, 
a physics-encoded FNO (PeFNO) is developed for a heterogeneous polycrystalline 
material consisting of isotropic elastic grains and subject to uniaxial extension. 
Stress field data for this purpose are obtained from the numerical 
solution of corresponding boundary-value problems for quasi-static 
mechanical equilibrium. 
For comparison with the PeFNO, this data is also employed to develop 
an analogous physics-guided FNO (PgFNO) and physics-informed FNO (PiFNO). 
As expected theoretically, and confirmed by this computational comparison, 
for comparable accuracy of the stress field itself as compared to the data, 
the stress field output by the trained and tested PeFNO is significantly more 
accurate in satisfying mechanical equilibrium than the output of either the 
PgFNO or the PiFNO. 

\end{abstract}

\begin{keyword} 
Scientific machine learning, 
physics-constrained neural approximations, 
divergence-free stress, 
heterogeneous solids
\end{keyword}

\end{frontmatter}

\section{Introduction}
\label{sec:intro}

Given the inherent lack and sparsity of data for most physical phenomena 
as well as their complexity, the development of empirical (i.e., data-based) 
or "surrogate" numerical models for these phenomena based on (artificial) 
neural networks (NNs) often includes constraints from physics to improve 
model robustness. As discussed recently by \cite{Faroughi2024} for example, 
the resulting physics-constrained NNs are currently of three 
types:~(i) physics-guided (PgNNs), 
(ii) physics-informed (PiNNs), 
and (iii) physics-encoded (PeNNs). 
For all three types, the data are constrained to be physical 
(e.g., experimental measurements); 
PiNNs and PeNNs are based on further physical constraints. 
In the case of PiNNs, these constraints are incorporated into the 
loss function (for training/testing), whereas for PeNNs, they are 
incorporated into ("encoded" in) the network architecture. 
By far the most common of these three types are PgNNs and PiNNs. 
PiNNs in particular have been developed for various applications in 
science and engineering \citep{cuomo2022scientific} such as 
computational fluid dynamics \citep{cai2021physics, mahmoudabadbozchelou2022nn} 
or heat transfer \citep{cai2021physics-Heat, xu2023physics, oommen2022solving}. 
Additional examples of PgNNs and PiNNs for computational fluid flow 
are discussed by \citet[][Tables 2, 4, 5]{Faroughi2024}. 
Analogously, a number of PgNNs and PiNNs have appeared 
for data-based numerical modeling in computational solid mechanics 
\citep{goswami2020transfer, abueidda2021meshless, 
mianroodi2021teaching, kumar2022machine, 
wessels2022computational, bai2023physics, diao2023solving, 
khorrami2023artificial, yang2023machine}; see also 
\citet[][Tables 3, 6, 7]{Faroughi2024}. 
Most recently, \cite{Heijden2025} have introduced a PiNN for 2D linear elasticity 
in which the constraint of mechanical equilibrium in the loss function is based on 
the Airy stress potential rather than on the divergence of the stress field.

Among the types of physics-constrained NNs mentioned above, 
PeNNs are the most challenging to develop \cite[e.g.,][\S 4]{Faroughi2024} 
and consequently the least common in the literature. Advantages 
of PeNNs include less sensitivity to data sparsity than exhibited 
by PgNNs and PiNNs. Prominent examples of PeNNs include 
(i) the physics-encoded recurrent convolutional neural network (PeRCNN) 
(ii) neural ordinary differential equations (NeuralODE), 
and (iii) neural partial differential equations (NeuralPDE) 
\citep{Dulny2021}. In particular, the latter represents an 
extension of NeuralODE to PDEs by combining the method of lines and 
NeuralODE in a multilayer convolutional NN. 
Being dependent on the method of lines, NeuralPDE is not directly 
applicable to certain kinds of PDEs (e.g., elliptical second-order PDEs). 

In the realm of computational solid mechanics, PeNNs are being developed 
for constitutive relations and balance relations. In contrast to balance relations, 
which apply to all materials, recall that constitutive relations hold only for 
a specific type of material behavior such as 
elastic, viscoelastic, elastic-viscoplastic, and so on. 
An example of these is the so-called constitutive artificial NN (CANN) 
of \cite{Linka2023} for isotropic elastic solids. 
More specifically, CANN represents a constitutive PeNN for finite-deformation 
isotropic hyperelasticity which exploits tensor invariants 
and encodes physically relevant mathematical 
properties such as polyconvexity in the network architecture. 
Direct generalization of CANNs to anisotropic and inelastic behavior 
\cite[e.g., iCANNs:][]{Holthusen2024} is possible 
with the help of network parameterization based for example 
on structure tensors \cite[e.g.,][]{Svendsen1994,Zheng1994Review}. 
Additional PeNNs (e.g., ICNNs) for scalar-valued constitutive relations 
like the stored energy, yield functions, or dual dissipation potentials, 
are discussed by \cite{Fuhg2025} in their recent review of data-driven 
constitutive relations for solids \cite[see also for example][]{Holthusen2026}.  

Rather than being based on functions as in the case of NNs, 
the approximation represented by so-called neural operators 
\cite[NOs:~e.g.,][\S 5]{Faroughi2024} is based on operators. 
Advantages of NOs over NNs include for example 
much less sensitivity to issues related to discretization. 
Prominent examples of NOs include deep operator networks 
\cite[DeepONets:~][]{Lu2019,Lu2021} and Fourier neural operators 
\cite[FNOs:~][]{Li2021,Kovachki2023}. PgNOs and PiNOs have been 
developed for data-based numerical modeling in computational solid 
mechanics, including heterogeneous materials \citep{kapoor2022comparison}, 
linear elastic fracture mechanics 
\citep{goswami2022fracture, goswami2023physics}, or even 
production by three-dimensional printing \citep{rashid2022learning}. 

The purpose of the current work is the development of an approach to 
encode quasi-static equilibrium (i.e., divergence-free) stress fields in 
neural approximations (NAs), which 
include both neural networks (NNs) and neural operators (NOs), 
in particular Fourier NOs (FNOs). To this end, the representation of 
such stress fields based on a stress potential is employed. 
To our knowledge, neither a PeNN, nor a PeNO, of this type for 
empirical, i.e., data-based numerical modeling of the quasi-static 
equilibrium stress field has appeared in the literature up to this point. 
Methodolgically close to the current work is that of \cite{RichterPowell2022}, 
who developed a PeNN for divergence-free vector fields 
(e.g., the velocity field in the case of incompressible flow) 
in computational fluid dynamics. 
In contrast to the methodology developed in the current work, 
their very interesting approach 
(i) is limited to vector fields, 
(ii) is based on the Hodge decomposition of differential 1-forms, 
and (iii) employs function approximation (i.e., NNs) and 
automatic differentiation. 
More recently, \cite{Jnini2025} have introduced a mixed PeNN-PiNN 
approach for computational fluid dynamics and magnetohydrodynamics 
in which mass and momentum balance are encoded in a so-called 
Riemann tensor neural network (RTNN) for symmetric divergence-free 
tensor 
fields. 
Although not explicitly pursued by them, their approach is potentially 
applicable to linear elastic solid dynamics as well. Also not pursued by 
\cite{Jnini2025} is the fact that the approach to divergence-free tensor 
fields underlying RTNNs can be generalized to non-symmetric stress tensors, 
as shown in \ref{app:RTNN} of the current work. 

The paper is organized as follows. 
Application of the Helmholtz decomposition to stress measures 
and quasi-static mechanical equilibrium is documented in 
Section \ref{sec:PotStsFPK}. For solids, the former include the 
non-linear first Piola-Kirchhoff (PK) stress in the paper as well 
as the linear symmetric stress. 
Fourier series forms of the corresponding potential relations and 
other relations relevant to the development of the PeFNO are 
also given in Section \ref{sec:PotStsFPK}. 
Section \ref{sec:PcNA} begins with a treatment and discussion 
of the architecture of physics-constrained NAs for the quasi-static 
equilibrium stress field in solids. 
In this context, the current approach to encode quasi-static mechanical 
equilibrium (i.e., divergence-free stress) in the NA architecture via a 
stress potential is treated in detail. 
To demonstrate the capabilities of the resulting PeFNO for divergence-free 
stress, it is applied in Section \ref{sec:ExaCom} to the stress field 
in a polycrystalline solid and compared with analogous Pg- and PiFNOs. 
Synthetic stress field data for training is generated via numerical solution 
of a physical boundary value problem (BVP) for quasi-static mechanical 
equilibrium in polycrystals consisting of isotropic elastic grains subject to 
uniaxial extension. 
The paper ends in Section \ref{sec:OutSum} with a summary and outlook. 
For completeness, a brief summary and comparison of theoretical aspects of 
the current approach as well as that of \cite{RichterPowell2022} are given 
in \ref{app:DecHH}. Likewise, the recent Riemann tensor NN  
(RTNN) approach of \cite{Jnini2025} for divergence-free symmetric tensor 
fields is briefly summarized in \ref{app:RTNN}. As mentioned above, 
although not considered by \cite{Jnini2025}, one can in fact generalize their 
approach to divergence-free non-symmetric tensor fields; this is also done 
in \ref{app:RTNN}. 

In the current work, 
three-dimensional Euclidean points or vectors are symbolized by lower case 
bold italic letters \(\bm{a},\ldots,\bm{z}\), and second-order Euclidean tensors 
by upper case bold italic letters \(\bm{A},\ldots,\bm{Z}\). In particular, let 
\(\bm{i}_{1},\bm{i}_{2},\bm{i}_{3}\) represent the Cartesian basis 
vectors, and \(\bm{I}\) the identity. The notation 
\(\bm{a}\cdot\bm{b}=a_{i}b_{i}\) and 
\(\bm{A}\cdot\bm{B}=A_{ij}B_{ij}\) (summation convention) 
denotes scalar products. 
The transpose \(\bm{A}^{\!\mathrm{T}}\) of any \(\bm{A}\), i.e., 
\(\bm{A}^{\!\mathrm{T}}\bm{a}\cdot\bm{b}:=\bm{a}\cdot\bm{A}\bm{b}\), 
determines its symmetric \(\mathop{\mathrm{sym}}\bm{A}
:=\tfrac{1}{2}(\bm{A}+\bm{A}_{}^{\mathrm{T}})\) and skew-symmetric 
\(\mathop{\mathrm{skw}}\bm{A}
:=\tfrac{1}{2}(\bm{A}-\bm{A}_{}^{\mathrm{T}})\) parts as usual. 
The tensor product \(\bm{a}\otimes\bm{b}\) is defined by 
\((\bm{a}\otimes\bm{b})\bm{c}:=(\bm{b}\cdot\bm{c})\,\bm{a}\); this 
determines 
\(\bm{a}\vee\bm{b}:=2\mathop{\mathrm{sym}}\bm{a}\otimes\bm{b}\) 
and \(\bm{a}\wedge\bm{b}:=2\mathop{\mathrm{skw}}\bm{a}\otimes\bm{b}\). 
Lastly, the axial tensor \(\mathop{\mathrm{axt}}\bm{a}\) of any vector 
\(\bm{a}\) is defined by 
\((\mathop{\mathrm{axt}}\bm{a})\,\bm{b}:=\bm{a}\times\bm{b}\). 
Additional concepts and notation are introduced as needed. 

\section{Mechanical equilbrium based on a stress potential}
\label{sec:PotStsFPK}

\subsection{Potential relations} 

The current FNO is developed to approximate stress fields in 
solids satisfying quasi-static linear and angular momentum 
balance \cite[e.g.,][]{Tru65,Silhavy1997}. In the geometrically 
non-linear (i.e., finite deformation) case, these balance relations 
take the forms\footnote{To be more precise, 
\eqref{equ:BalMomQuaStaNonGeo} represent the "referential" or "Lagrangian" 
forms of quasi-static linear and angular momentum balance 
for so-called "non-polar" solid materials.} 
\begin{equation}
\mathop{\mathrm{div}}\bm{P}=\bm{0}
\,,\quad
\bm{F}\bm{P}^{\mathrm{T}}
=\bm{P}\bm{F}^{\mathrm{T}}
\,,
\label{equ:BalMomQuaStaNonGeo}
\end{equation}
respectively, where \(\bm{P}\) and \(\bm{F}\) are the first Piola-Kirchhoff 
stress and deformation gradient, respectively. On the other hand, in the 
geometrically linear (i.e., infinitesimal deformation) case, these are given 
by\footnote{Relations of the form \eqref{equ:BalMomQuaStaLinGeo} also 
hold in the quasi-static, geometrically non-linear case for the Cauchy stress 
with respect to the current ("Eulerian") configuration. For general non-linear 
solid mechanics, however, \eqref{equ:BalMomQuaStaNonGeo} are relevant.}  
\begin{equation}
\mathop{\mathrm{div}}\bm{T}
=\bm{0}
\,,\quad
\bm{T}^{\mathrm{T}}=\bm{T}
\,,
\label{equ:BalMomQuaStaLinGeo}
\end{equation}
with \(\bm{T}\) the corresponding stress measure. In contrast to \(\bm{P}\) 
in the non-linear case, then, the stress measure \(\bm{T}\) is symmetric 
in the linear case. 
For the current class of (non-polar) solids, note that angular momentum 
balance in either form \eqref{equ:BalMomQuaStaNonGeo}${}_{2}$ or 
\eqref{equ:BalMomQuaStaLinGeo}${}_{2}$ is satisfied identically by 
the constitutive relation for the stress. 

The stress potential relations in the current work are based on a generalization 
of the Helmholtz-Hodge decomposition of vector fields \cite[e.g.,][]{Bhatia2013} 
to second-order tensor fields as discussed in more detail in \ref{app:DecHH}. In 
particular, the potential relation 
\begin{equation}
\bm{P}=\mathop{\mathrm{curl}}\bm{A}
\label{equ:PotStsFPK}
\end{equation}
for \(\bm{P}\) is obtained directly from \eqref{equ:FieTenGenFreDiv} via 
the choices \(\bm{S}=\bm{P}\) and \(\bm{\Phi}=\bm{A}\) and satisfies 
\eqref{equ:BalMomQuaStaNonGeo}${}_{1}$ identically. In contrast 
to \(\bm{P}\), \eqref{equ:FieTenGenFreDiv} is not directly applicable to 
\(\bm{T}\) since it does not preserve the symmetry of \(\bm{T}\). 
Rather than by \eqref{equ:FieTenGenFreDiv}, then, the potential relation for 
\(\bm{T}\) is given by 
\begin{equation}
\bm{T}
=\mathop{\mathrm{inc}}\bm{B}
:=\mathop{\mathrm{curl}}
\,(\mathop{\mathrm{curl}}\bm{B})^{\mathrm{T}}
\label{equ:PotStsBel}
\end{equation} 
with the corresponding potential \(\bm{B}\) being 
symmetric. In the context of 
\eqref{equ:PotStsBel}, the symmetry of \(\bm{B}\) results in \(\bm{T}\) 
satisfying angular momentum balance \eqref{equ:BalMomQuaStaLinGeo}${}_{2}$ 
identically (as evident for example from the algebraic relation 
\eqref{equ:PotStsCofFor}${}_{2}$ below);  
likewise, \(\bm{T}\) 
then satisfies linear momentum balance 
\eqref{equ:BalMomQuaStaLinGeo}${}_{1}$ identically. 
In linear elasticity, the operator \(\mathop{\mathrm{inc}}\) is known 
as the incompatibility operator \cite[e.g.,][]{Teo82}, and 
\(\bm{B}\) is known as the Beltrami stress potential or "function"  
\cite[e.g.,][\S13.6]{Sadd2009}. 

In PeNNs, operator relations like \eqref{equ:PotStsFPK} or \eqref{equ:PotStsBel}
are evaluated in discretized form, for example via automatic differentiation 
given a differentiable architecture \cite[e.g.,][]{RichterPowell2022}. 
In the PeFNO developed in the current work, these are evaluated 
with the help of Fourier methods and the corresponding discretized 
algebraic relations in Fourier space as documented in what follows. 

\subsection{Fourier series representations} 

Any integrable field \(f\) on some region \(U\) of three-dimensional Euclidean space 
can be expressed as the sum 
\begin{equation}
\textstyle
f(\bm{x})
=\bar{f}+\tilde{f}(\bm{x})
\,,\quad
\bar{f}
:=\frac{1}{v(U)}
\int_{U}f(\bm{x})\ dv(\bm{x})
\,,\ \ 
\tilde{f}(\bm{x})
:=f(\bm{x})-\bar{f}
\,,
\label{equ:SplFluMea}
\end{equation} 
of its (spatial constant) 
mean \(\bar{f}\) and fluctuation \(\tilde{f}\) parts on \(U\). Applying this 
for example to the first Piola-Kirchhoff stress \(\bm{P}\), one obtains the 
reduced form 
\begin{equation}
\mathop{\mathrm{div}}\bm{P}
=\mathop{\mathrm{div}}\tilde{\bm{P}}
=\bm{0}
\label{equ:RedBalMomQuaStaNonGeo}
\end{equation}
of \eqref{equ:BalMomQuaStaNonGeo}${}_{1}$, and so that 
\begin{equation}
\bar{\bm{P}}=\bar{\bm{A}}
\,,\quad
\tilde{\bm{P}}
=\mathop{\mathrm{curl}}\bm{A}
=\mathop{\mathrm{curl}}\tilde{\bm{A}}
\label{equ:RedPotStsFPK}
\end{equation}
of the corresponding potential relation \eqref{equ:PotStsFPK}. The analogous 
reduced form 
\begin{equation}
\bar{\bm{T}}=\bar{\bm{B}}
\,,\quad
\tilde{\bm{T}}
=\mathop{\mathrm{inc}}\bm{B}
=\mathop{\mathrm{inc}}\tilde{\bm{B}}
\label{equ:RedPotStsBel}
\end{equation} 
of \eqref{equ:PotStsBel} for the geometric linear case then holds as well. 

Assume next that the stress and potential fields are periodic on \(U\). 
The Fourier series of any field \(f\) integrable and periodic on \(U\) is given by 
\begin{equation}
\textstyle
f(\bm{x})
=\sum_{\bm{k}\in U^{\ast}}
e_{}^{\imath{}\bm{k}\cdot\bm{x}}
\hat{f}(\bm{k})
\,,\quad
\hat{f}(\bm{k})
=\frac{1}{v(U)}
\int_{U}
e_{}^{-\imath{}\bm{k}\cdot\bm{x}}
\,f(\bm{x})
\ dv(\bm{x})
\,,
\label{equ:FieScaSerFor}
\end{equation} 
where \(\imath=\sqrt{-1}\), \(U^{\ast}\) is the reciprocal (wavevector)  
space of \(U\), and \(v(U)=\int_{U}dv(\bm{x})\) is the volume of \(U\). 
In the context of \eqref{equ:FieScaSerFor}, note that 
\begin{equation}
\textstyle
\bar{f}
=\hat{f}(\bm{0})
\,,\ \ 
\tilde{f}(\bm{x})
=\sum_{\bm{k}\neq\bm{0}}
e_{}^{\imath{}\bm{k}\cdot\bm{x}}
\hat{f}(\bm{k})
\,,
\label{equ:SplFluMeaSerFou}
\end{equation}
from \eqref{equ:SplFluMea} with 
\(\sum_{\bm{k}\neq\bm{0}}
:=\sum_{\bm{k}\in U^{\ast}\setminus\lbrace\bm{0}\rbrace}\). 
In particular, then, one obtains the forms 
\begin{equation}
\begin{array}{rclcl}
\bm{P}(\bm{x})
&=&
\hat{\bm{P}}(\bm{0})
+\sum_{\bm{k}\neq\bm{0}}
e^{\imath\bm{k}\cdot\bm{x}}
\,\hat{\bm{P}}(\bm{k})
&=&
\bar{\bm{P}}+\tilde{\bm{P}}(\bm{x})
\,,\\
\bm{T}(\bm{x})
&=&
\hat{\bm{T}}(\bm{0})
+\sum_{\bm{k}\neq\bm{0}}
e^{\imath\bm{k}\cdot\bm{x}}
\,\hat{\bm{T}}(\bm{k})
&=&
\bar{\bm{T}}+\tilde{\bm{T}}(\bm{x})
\,,\\
\bm{A}(\bm{x})
&=&
\hat{\bm{A}}(\bm{0})
+\sum_{\bm{k}\neq\bm{0}}
e^{\imath\bm{k}\cdot\bm{x}}
\,\hat{\bm{A}}(\bm{k})
&=&
\bar{\bm{A}}+\tilde{\bm{A}}(\bm{x})
\,,\\
\bm{B}(\bm{x})
&=&
\hat{\bm{B}}(\bm{0})
+\sum_{\mathbf{k}\neq\bm{0}}
e^{\imath\mathbf{k}\cdot\mathbf{x}}
\,\hat{\bm{B}}(\bm{k})
&=&
\bar{\bm{B}}+\tilde{\bm{B}}(\bm{x})
\,,
\end{array}
\label{equ:StsPotSerForFluMea}
\end{equation} 
for the stress and potential fields, and so the operator relations 
\begin{equation}
\begin{array}{rclcl}
\mathop{\mathrm{div}}\bm{P}(\bm{x})
&=&
\mathop{\mathrm{div}}\skew5\tilde{\bm{P}}(\bm{x})
&=&
\sum_{\bm{k}\neq\bm{0}}
e_{}^{\imath{}\bm{k}\cdot\bm{x}}
\skew5\hat{\bm{P}}(\bm{k})
\,\imath\bm{k}
\,,\\
\mathop{\mathrm{div}}\bm{T}(\bm{x})
&=&
\mathop{\mathrm{div}}\skew2\tilde{\bm{T}}(\bm{x})
&=&
\sum_{\bm{k}\neq\bm{0}}
e_{}^{\imath{}\bm{k}\cdot\bm{x}}
\skew2\hat{\bm{T}}(\bm{k})
\,\imath\bm{k}
\,,\\
\mathop{\mathrm{curl}}\bm{A}(\bm{x})
&=&
\mathop{\mathrm{curl}}\skew5\tilde{\bm{A}}(\bm{x})
&=&
\sum_{\bm{k}\neq\bm{0}}
e_{}^{\imath{}\bm{k}\cdot\bm{x}}
\skew5\hat{\bm{A}}(\bm{k})
\,(\mathop{\mathrm{axt}}\imath\bm{k})^{\mathrm{T}}
\,,\\
\mathop{\mathrm{inc}}\bm{B}(\bm{x})
&=&
\mathop{\mathrm{inc}}\skew4\tilde{\bm{B}}(\bm{x})
&=&
\sum_{\bm{k}\neq\bm{0}}
e_{}^{\imath{}\bm{k}\cdot\bm{x}}
(\mathop{\mathrm{axt}}\imath\bm{k})
\,\skew4\hat{\bm{B}}(\bm{k})
\,(\mathop{\mathrm{axt}}\imath\bm{k})^{\mathrm{T}}
\,.
\end{array}
\label{equ:StsKirPioFirPotSerForCur}
\end{equation} 
Together, \eqref{equ:StsPotSerForFluMea} and  
\eqref{equ:StsKirPioFirPotSerForCur}${}_{3,4}$ 
determine the corresponding Fourier series component relations 
\begin{equation}
\begin{array}{rcl}
\skew3\hat{\bm{P}}(\bm{k})
&=&
\left\lbrace
\begin{array}{lcl}
\skew5\hat{\bm{A}}(\bm{k})
&&
\bm{k}=\bm{0}
\\
\skew5\hat{\bm{A}}(\bm{k})
\,(\mathop{\mathrm{axt}}\imath\bm{k})^{\mathrm{T}}
&&
\bm{k}\neq\bm{0}
\end{array}
\right.
\,,
\\[5mm]
\skew3\hat{\bm{T}}(\bm{k})
&=&
\left\lbrace
\begin{array}{lcl}
\skew4\hat{\bm{B}}(\bm{k})
&&
\bm{k}=\bm{0}
\\
(\mathop{\mathrm{axt}}\imath\bm{k})
\,\skew4\hat{\bm{B}}(\bm{k})
\,(\mathop{\mathrm{axt}}\imath\bm{k})^{\mathrm{T}}
&&
\bm{k}\neq\bm{0}
\end{array}
\right.
\,,
\end{array}
\label{equ:PotStsCofFor}
\end{equation} 
in the context of \eqref{equ:RedPotStsFPK} and \eqref{equ:RedPotStsBel}, 
respectively (recall that \(\bm{B}\) is symmetric). 

As treated in more detail 
in the sequel, the relations \eqref{equ:PotStsCofFor} between Fourier 
coefficients of the stress and corresponding potentials are employed to 
encode mechanical equilibrium in the architecture of NAs for the stress field. 
In the rest of the work, attention is focused for brevity on 
the non-linear case based on \eqref{equ:BalMomQuaStaNonGeo} and 
\eqref{equ:PotStsFPK}. Treatment of the linear case based on 
\eqref{equ:BalMomQuaStaLinGeo} and \eqref{equ:PotStsBel} is analogous. 

\subsection{Cartesian component relations}

Being computational in character, 
the NAs and in particular the PeFNO are based on the Cartesian component 
forms of the above Euclidean tensor relations. 
In particular, \(P_{\!i\!j}:=\bm{i}_{i}\cdot\bm{P}\bm{i}_{\!j}\) 
and \(A_{i\!j}:=\bm{i}_{i}\cdot\bm{A}\bm{i}_{\!j}\) determine 
the component matrix forms 
\begin{equation}
\textstyle
\mathbf{P}(\mathbf{x})
=\sum_{\mathbf{k}\in\mathrm{U}^{\ast}}
e^{\imath\mathbf{k}\cdot\mathbf{x}}
\,\hat{\mathbf{P}}(\mathbf{k})
\,,\ \ 
\mathbf{A}(\mathbf{x})
=\sum_{\mathbf{k}\in\mathrm{U}^{\ast}}
e^{\imath\mathbf{k}\cdot\mathbf{x}}
\hat{\mathbf{A}}(\mathbf{k})
\,,
\label{equ:PotStsRepFouMatCar}
\end{equation} 
on \(\mathrm{U}\subset\mathbb{R}^{3}\) for the Fourier series 
of \(\bm{P}\) and \(\bm{A}\), respectively. Likewise, 
\begin{equation}
\textstyle
\hat{\mathbf{P}}(\mathbf{k})
=\left\lbrace
\begin{array}{lcl}
\hat{\mathbf{A}}(\mathbf{k})
&&
\mathbf{k}=\bm{0}
\\
\hat{\mathbf{A}}(\mathbf{k})
\,(\mathop{\mathrm{axt}}\imath\mathbf{k})^{\mathrm{T}}
&&
\mathbf{k}\neq\bm{0}
\end{array}
\right.
,
\quad
\mathop{\mathrm{axt}}\imath\mathbf{k}
=\imath
\left\lbrack
\begin{array}{ccc}
0&-k_{3}&k_{2}
\\
k_{3}&0&-k_{1}
\\
-k_{2}&k_{1}&0
\end{array}
\right\rbrack
,
\label{equ:PotStsRepFouCof}
\end{equation} 
holds for the component matrix forms of \eqref{equ:PotStsCofFor}${}_{1}$ 
and \(\mathop{\mathrm{axt}}\imath\mathbf{k}\), respectively, with 
\begin{equation}
\begin{array}{l}
\hat{\mathbf{A}}(\mathbf{k})
\,(\mathop{\mathrm{axt}}\imath\mathbf{k})^{\mathrm{T}}
\\
\quad=\ \ \imath
\left\lbrack
\begin{array}{ccc}
k_{2}\hat{A}_{13}(\mathbf{k})-k_{3}\hat{A}_{12}(\mathbf{k})&
k_{3}\hat{A}_{11}(\mathbf{k})-k_{1}\hat{A}_{13}(\mathbf{k})&
k_{1}\hat{A}_{12}(\mathbf{k})-k_{2}\hat{A}_{11}(\mathbf{k})
\\
k_{2}\hat{A}_{23}(\mathbf{k})-k_{3}\hat{A}_{22}(\mathbf{k})&
k_{3}\hat{A}_{21}(\mathbf{k})-k_{1}\hat{A}_{23}(\mathbf{k})&
k_{1}\hat{A}_{22}(\mathbf{k})-k_{2}\hat{A}_{21}(\mathbf{k})
\\
k_{2}\hat{A}_{33}(\mathbf{k})-k_{3}\hat{A}_{32}(\mathbf{k})&
k_{3}\hat{A}_{31}(\mathbf{k})-k_{1}\hat{A}_{33}(\mathbf{k})&
k_{1}\hat{A}_{32}(\mathbf{k})-k_{2}\hat{A}_{31}(\mathbf{k})
\end{array}
\right\rbrack
\end{array}
\,.
\label{equ:PotStsRepFouCofMat}
\end{equation} 
Lastly, the component array form 
\begin{equation}
\textstyle
\mathbf{d}(\mathbf{x})
:=\mathop{\mathrm{div}}\tilde{\mathbf{P}}(\mathbf{x})
=\sum_{\mathbf{k}\neq\bm{0}}
e^{\imath\mathbf{k}\cdot\mathbf{x}}
\,\skew3\hat{\mathbf{d}}(\mathbf{k})
\label{equ:StsDivRepFouAryCar}
\end{equation} 
holds for \eqref{equ:StsKirPioFirPotSerForCur}${}_{1}$ with 
\begin{equation}
\skew3\hat{\mathbf{d}}(\mathbf{k})
:=\hat{\mathbf{P}}(\mathbf{k})
\,\imath\mathbf{k}
=\imath
\left\lbrack
\begin{array}{c}
\hat{P}_{11}(\mathbf{k})\,k_{1}
+\hat{P}_{12}(\mathbf{k})\,k_{2}
+\hat{P}_{13}(\mathbf{k})\,k_{3}
\\
\hat{P}_{21}(\mathbf{k})\,k_{1}
+\hat{P}_{22}(\mathbf{k})\,k_{2}
+\hat{P}_{23}(\mathbf{k})\,k_{3}
\\
\hat{P}_{31}(\mathbf{k})\,k_{1}
+\hat{P}_{32}(\mathbf{k})\,k_{2}
+\hat{P}_{33}(\mathbf{k})\,k_{3}
\end{array}
\right\rbrack
\,,\quad
\mathbf{k}\neq\mathbf{0}
\,.
\label{equ:ModStsDivRepFouAryCar}
\end{equation} 
In physical space, note that \(\mathbf{d}=(d_{1},d_{2},d_{3})\) with 
\(d_{i}:=\partial P_{\!i\!j}/\partial x_{\!j}\) (summation convention). 
As will be seen below, the PeNA output transformation is determined 
by \eqref{equ:PotStsRepFouCof}, and \eqref{equ:StsDivRepFouAryCar} 
is employed in the PiNA loss function.

\section{PcNAs for equilibrium stress fields}
\label{sec:PcNA}

In the current work, physics-constrained NAs and in particular PcFNOs 
are developed as empirical (i.e., data-based) models for quasi-static 
equilibrium stress fields in solids. Since such stress fields are the output 
of the NA for any form of NA input, the latter is represented generically  
in the NA architecture to begin with in the following. An example for the 
form of NA input for the case of data generated from a physical model 
is given at the end of this section. 

\subsection{NA architecture}

For simplicity, attention is restricted here to the case of a  
"multilayer" NA \cite[e.g.,][]{Kovachki2023}. 
Such an NA can be represented in the generic form
\begin{equation}
\mathbf{o}
=\bm{\tau}_{\mathrm{NA}}\diamond\mathbf{i}
\,,
\label{equ:LayMulAppNeu}
\end{equation} 
where \(\mathbf{i}(\mathbf{x})\) is a real-array-valued input field 
on \(\mathbb{R}^{3}\), and \(\mathbf{o}(\mathbf{x})\) a corresponding 
output field (e.g., an array of the Cartesian components of the stress field). 
In the current multilayer case, 
\(\bm{\tau}_{\mathrm{NA}}\) in \eqref{equ:LayMulAppNeu} 
is determined by generalized composition \(\diamond\) 
(e.g., standard function composition, or convolution) 
of (i) 1 input transformation \(\bm{\tau}_{\mathrm{inp}}\), 
(ii) \(n_{\mathrm{hid}}\) "hidden" transformations 
\(\bm{\tau}_{1},\ldots,\bm{\tau}_{n_{\mathrm{hid}}}\), 
and (iii) 1 output transformation \(\bm{\tau}_{\mathrm{out}}\), i.e., 
\begin{equation}
\bm{\tau}_{\mathrm{NA}}
=\bm{\tau}_{\mathrm{out}}
\diamond
\bm{\tau}_{\mathrm{hid}}
\diamond
\bm{\tau}_{\mathrm{inp}}
\,,\quad
\bm{\tau}_{\mathrm{hid}}
=\bm{\tau}_{n_{\mathrm{hid}}}
\diamond
\cdots
\diamond
\bm{\tau}_{1}
\,.
\label{equ:AppNeuLayMul}
\end{equation}
Here, 
\begin{equation}
\mathbf{h}_{0}
=\bm{\tau}_{\mathrm{inp}}\diamond\mathbf{i}
\,,\quad
\mathbf{h}_{l}
=\bm{\tau}_{l}\diamond\mathbf{h}_{l-1}
\,,\ \ 
l=1,\ldots,n_{\mathrm{hid}}
\,,\quad
\mathbf{o}
=\bm{\tau}_{\mathrm{out}}\diamond\mathbf{h}_{n_{\mathrm{hid}}}
\,,
\label{equ:OutHidInAppNeuLayMul}
\end{equation}
with \(\mathbf{h}_{l}\) the (real-array-valued) hidden field in layer \(l\). 
As usual, each hidden transformation \(\bm{\tau}_{l}\) in 
\eqref{equ:AppNeuLayMul}${}_{2}$ is given by the composition 
\begin{equation}
\bm{\tau}_{l}
=\bm{\sigma}_{\!l}\circ\bm{\alpha}_{l}
\,,\quad
\mathbf{a}_{l}
=\bm{\alpha}_{l}\diamond\mathbf{h}_{l-1}
\,,\quad
\mathbf{h}_{l}
=\bm{\sigma}_{\!l}(\mathbf{a}_{l})
:=(\sigma(a_{l1}),\sigma(a_{l2}),\ldots)
\,,
\label{equ:TraLayHidGen}
\end{equation}
of affine \(\bm{\alpha}_{l}\) and activation \(\bm{\sigma}_{\!l}\) 
parts, the latter determined by the non-linear ("neuron") activation 
function \(\sigma\). The affine transformation \(\bm{\alpha}_{l}\) 
is defined by 
\begin{equation}
\begin{array}{rclcl}
\bm{\alpha}_{l}\diamond\mathbf{h}_{l-1}
&=&
\bm{\alpha}^{\mathrm{NN}}(\mathbf{W}_{\!l},\mathbf{b}_{l})
\diamond
\mathbf{h}_{l-1}
&:=&
\mathbf{W}_{\!l}\mathbf{h}_{l-1}
+\mathbf{b}_{l}
\,,\\
\bm{\alpha}_{l}\diamond\mathbf{h}_{l-1}
&=&
\bm{\alpha}^{\mathrm{NO}}
(\mathbf{O}_{l},\mathbf{W}_{\!l},\mathbf{b}_{l})
\diamond
\mathbf{h}_{l-1}
&:=&
\mathbf{O}_{l}\diamond\mathbf{h}_{l-1}
+\mathbf{W}_{\!l}\mathbf{h}_{l-1}
+\mathbf{b}_{l}
\,,
\end{array}
\label{equ:TraLayHidNA}
\end{equation}
for NNs and NOs, respectively, where 
\(\mathbf{W}_{\!l}\) is the weight matrix, 
\(\mathbf{b}_{l}\) the bias array, and 
\(\mathbf{O}_{l}\) the operator part of the NO. 
Formally, \(\bm{\alpha}^{\mathrm{NN}}\) is clearly 
a special case of \(\bm{\alpha}^{\mathrm{NO}}\). 
Of particular interest here are FNOs \cite[e.g.,][]{Li2021,Kovachki2023}, 
for which \(\mathbf{O}_{l}\) in \eqref{equ:TraLayHidNA}${}_{2}$ 
takes the form 
\begin{equation}
\begin{array}{rclcl}
\mathbf{O}_{l}\diamond\mathbf{h}_{l-1}
=\mathbf{O}^{\mathrm{FNO}}(\mathbf{K}_{l})
\diamond\mathbf{h}_{l-1}
&:=&
\mathcal{F}^{-1}
\lbrack
\mathcal{F}\lbrack\mathbf{K}_{l}\rbrack
\,\mathcal{F}\lbrack\mathbf{h}_{l-1}\rbrack\rbrack
&=&
\mathcal{F}^{-1}
\lbrack\hat{\mathbf{K}}_{l}\!\hat{\mathbf{\,h}}_{l-1}\rbrack
\end{array}
\label{equ:TraOpeLayHidFNO}
\end{equation} 
in terms of the Fourier transform \(\mathcal{F}\) 
and kernel matrix field \(\mathbf{K}_{l}\). 
In contrast to the hidden case and \eqref{equ:TraLayHidGen}, 
the input \eqref{equ:OutHidInAppNeuLayMul}${}_{1}$ 
and output \eqref{equ:OutHidInAppNeuLayMul}${}_{3}$ 
transformations involve no activation. 
These are treated in more detail in the sequel. 

So-called hyperparameters of the NA based on 
\eqref{equ:LayMulAppNeu}-\eqref{equ:TraLayHidNA} 
include 
(i) the number of neurons per layer \(n_{l}^{\mathrm{neu}}\) (the "width" of \(l\)), 
(ii) the activation function \(\sigma\), and 
(iii) the number of hidden layers \(n_{\mathrm{hid}}\) 
(the "depth" of the NA). 
Assuming that \(\mathbf{i}\) and \(\mathbf{h}_{l}\) 
take values in \(\mathbb{R}^{d(\mathbf{i})}\) and 
\(\mathbb{R}^{d(\mathbf{h}_{l})}\), respectively, note for example that 
\(d(\mathbf{h}_{l})=n_{l}^{\mathrm{neu}}d(\mathbf{i})\) holds. 
These parameters are discussed further in what follows. 

\subsection{Output transformations for the equilibrium stress field}

The basic difference between the PeNA and the other two (i.e., PgNA and PiNA) 
for the equilibrium stress field lies in the form of the output 
transformation \eqref{equ:OutHidInAppNeuLayMul}${}_{3}$, i.e., 
\(\mathbf{o}
=\bm{\tau}_{\mathrm{out}}\diamond\mathbf{h}_{n_{\mathrm{hid}}}\). 
For both cases, \(\mathbf{o}=\mathbf{p}^{\mathrm{out}}\) in terms of 
the notation \(\mathbf{p}:=(\mathbf{p}_{1},\mathbf{p}_{2},\mathbf{p}_{3})
\equiv(P_{\!11},P_{\!12},\ldots,P_{\!32},P_{\!33})
\), where \(\mathbf{p}_{i}:=(P_{\!i1},P_{\!i2},P_{\!i3})\) 
is the \(i^{\mathrm{th}}\) row of \(\mathbf{P}\). Employing the same row 
notation \(\mathbf{a}:=(\mathbf{a}_{1},\mathbf{a}_{2},\mathbf{a}_{3})
\equiv(A_{11},A_{12},\ldots,A_{32},A_{33})\) 
with \(\mathbf{a}_{i}:=(A_{i1},A_{i2},A_{i3})\) for \(\mathbf{A}\), 
as well as the auxiliary field 
\begin{equation}
\textstyle
\mathbf{s}
:=\left\lbrace
\begin{array}{lcl}
\mathbf{p}
&&
\hbox{PgNA, PiNA}
\\
\mathbf{a}
&&
\hbox{PeNA}
\end{array}
\right.
\,,
\label{equ:TraOutAuxPcFNO}
\end{equation} 
\(\bm{\tau}_{\mathrm{out}}\) is determined by
\begin{equation}
\textstyle
\mathbf{p}^{\mathrm{out}}(\mathbf{x})
=\hat{\mathbf{s}}^{\mathrm{out}}(\bm{0})
+\sum_{\mathbf{k}\neq\bm{0}}
e^{\imath\mathbf{k}\cdot\mathbf{x}}
\left\lbrace
\begin{array}{lcl}
\hat{\mathbf{s}}^{\mathrm{out}}(\mathbf{k})
&&
\hbox{PgNA, PiNA}
\\
\imath\mathbf{k}
\times
\hat{\mathbf{s}}^{\mathrm{out}}(\mathbf{k})
&&
\hbox{PeNA}
\end{array}
\right.
\label{equ:TraOutPcFNO}
\end{equation} 
for the current PcNAs with 
\begin{equation}
\hat{\mathbf{s}}^{\mathrm{out}}(\mathbf{k})
:=\left\lbrace
\begin{array}{lcl}
\mathbf{W}_{\!\mathrm{out}}
\hat{\,\mathbf{h}}_{n_{\mathrm{hid}}}(\mathbf{k})
+\mathbf{b}_{\mathrm{out}}
&&
\mathrm{NN}
\\
\lbrack
\hat{\mathbf{K}}_{\mathrm{out}}(\mathbf{k})
+\mathbf{W}_{\!\mathrm{out}}
\rbrack
\hat{\,\mathbf{h}}_{n_{\mathrm{hid}}}(\mathbf{k})
+\mathbf{b}_{\mathrm{out}}
&&
\mathrm{FNO}
\end{array}
\right.
\label{equ:HidOutcFNO}
\end{equation} 
from \eqref{equ:TraLayHidNA} and \eqref{equ:TraOpeLayHidFNO}. 
The notation 
\(\imath\mathbf{k}\times\hat{\mathbf{s}}
\equiv\imath\mathbf{k}\times\hat{\mathbf{a}}
:=(\imath\mathbf{k}\times\hat{\mathbf{a}}_{1},
\imath\mathbf{k}\times\hat{\mathbf{a}}_{2},
\imath\mathbf{k}\times\hat{\mathbf{a}}_{3})\) 
in \eqref{equ:TraOutPcFNO} is based on 
\(\hat{\mathbf{p}}_{i}=\imath\mathbf{k}\times\hat{\mathbf{a}}_{i}\) 
from \eqref{equ:PotStsRepFouCof}. 

For the purposes of checking 
\(\mathbf{d}^{\mathrm{out}}
=\mathop{\mathrm{div}}\mathbf{P}^{\mathrm{out}}\), 
as well as for PiNA training and testing, \eqref{equ:TraOutPcFNO} can be extended to 
\((\mathbf{p}^{\mathrm{out}}, 
\mathbf{d}^{\mathrm{out}})\) via \eqref{equ:StsDivRepFouAryCar} with 
\begin{equation}
\textstyle
d_{i}^{\mathrm{out}}(\mathbf{x})
=\sum_{\mathbf{k}\neq\bm{0}}
e^{\imath\mathbf{k}\cdot\mathbf{x}}
\left\lbrace
\begin{array}{lcl}
\imath\mathbf{k}\cdot\hat{\mathbf{s}}_{i}^{\mathrm{out}}(\mathbf{k})
&&
\hbox{PgNA, PiNA}
\\
\imath\mathbf{k}
\times
\imath\mathbf{k}
\cdot
\hat{\mathbf{s}}_{i}^{\mathrm{out}}(\mathbf{k})
&&
\hbox{PeNA}
\end{array}
\right.
\label{equ:TraDivOutPcFNO}
\end{equation} 
(recall \(\mathbf{d}:=(d_{1},d_{2},d_{3})\) and 
\(\mathbf{s}=(\mathbf{s}_{1},\mathbf{s}_{2},\mathbf{s}_{3})\) 
from \eqref{equ:TraOutAuxPcFNO}). 
Note that both 
\(\mathbf{p}^{\mathrm{out}}\) and \(\mathbf{d}^{\mathrm{out}}\) 
are determined by \(\hat{\mathbf{s}}^{\mathrm{out}}(\mathbf{k})\) 
from \eqref{equ:HidOutcFNO}. In the latter case, note also that 
\(\skew3\hat{\mathbf{d}}^{\mathrm{out}}(\mathbf{k})
=\hat{\mathbf{P}}^{\mathrm{out}}(\mathbf{k})\,\imath\mathbf{k}
=\hat{\mathbf{A}}^{\!\mathrm{out}}(\mathbf{k})
\,(\imath\mathbf{k}\times\imath\mathbf{k})=\bm{0}\) 
holds identically for the PeNA 
from \eqref{equ:PotStsRepFouCof} and \eqref{equ:ModStsDivRepFouAryCar}
for \(\mathbf{k}\neq\bm{0}\). 
Approximation errors due for example to truncation and discretization 
of the Fourier series relations result, however, in 
\(\mathbf{d}^{\mathrm{out}}\neq\bm{0}\), even for PeNA output. 
Nevertheless, the value of \(|\mathbf{d}^{\mathrm{out}}|\) 
for PeNAs is still orders-of-magnitude smaller than for either Pg- or PiNAs, 
as documented by the computational results to follow. 

\subsection{Example for NA input} 
\label{sec:InpNAExa}

When the data are generated from physical models based on the 
numerical solution of BVPs, data "labeling" and the form of NA input 
\(\mathbf{i}(\mathbf{x})\) are determined by the corresponding 
constitutive relations and boundary conditions of the physical model. 
As an example, consider the case of an elastic polycrystal in which 
each grain in the microstructure is modeled as an isotropic 
elastic solid via the Saint-Venant-Kirchhoff relation  
\begin{equation}
\mathbf{P}(E,\nu,\mathbf{F}) 
=\frac{E\nu}{(1+\nu)(1-2\nu)}
\,(\mathbf{I}\cdot\mathbf{E})\,\mathbf{F}
+\frac{E}{1+\nu}
\,\mathbf{F}\mathbf{E}
\,,
\label{equ:ElaKirVenSai}
\end{equation} 
in component matrix form. Here, \(E\) and \(\nu\) represent 
Young's modulus and Poisson's ratio, respectively, 
\(\mathbf{F}\) is the component matrix of the deformation gradient, and 
\begin{equation}
\mathbf{E}
:=\tfrac{1}{2}(\mathbf{F}^{\mathrm{T}}\mathbf{F}-\mathbf{I})
\label{equ:StnGreMatComCar}
\end{equation}
is the corresponding matrix of the symmetric Green strain.  
In addition, \(\mathbf{I}\cdot\mathbf{E}=E_{11}+E_{22}+E_{33}\) 
is the trace of \(\mathbf{E}\). 
Since the solids under consideration are materially heterogeneous, 
note that material properties like \(E\) and \(\nu\) are also fields on 
\(\mathrm{U}\). 
For a polycrystal in which each grain represents the same 
material, for example, varying \(E\) and \(\nu\) from grain 
to grain models the case in which each grain has a 
different orientation. 

In addition to \eqref{equ:ElaKirVenSai}, consider for example BVPs for 
mechanical equilbrium based on 
kinematic boundary conditions taking the form of prescribed values for 
the components \(\bar{\mathbf{F}}\) of the mean deformation gradient 
imposed on the unit cell \(\mathrm{U}\subset\mathbb{R}^{3}\). 
For \eqref{equ:ElaKirVenSai} and such boundary conditions, then, 
\begin{equation}
\mathbf{i}(\mathbf{x})
=(E(\mathbf{x}),\nu(\mathbf{x}),\bar{\mathbf{f}})
\label{equ:InpFNO}
\end{equation}
is the form taken by the NA input. Here, 
\(\mathbf{f}:=(\mathbf{f}_{1},\mathbf{f}_{2},\mathbf{f}_{3}) 
\equiv(F_{\!11},F_{\!12},\ldots,F_{\!32},F_{\!33})\), analogous to \(\mathbf{p}\) 
for the first PK stress above. 
Given \eqref{equ:InpFNO}, note that \(d(\mathbf{i})=11\) 
and \(d(\mathbf{h}_{l})=11\,n_{l}^{\mathrm{neu}}\) for the current 
example. 

\section{Computational examples}
\label{sec:ExaCom}

As an example application of the current approach to PeNAs for 
divergence-free stress, a PeFNO is developed in the following for data-based 
modeling of the quasi-static equilibrium stress field in a heterogeneous elastic 
solid. For comparison, analogous Pg- and PiFNOs are also developed. 

\subsection{Data generation}

Stress field data are generated via numerical solution of 
BVPs for quasi-static mechanical equilibrium discussed in 
Section \ref{sec:InpNAExa} corresponding to the form 
\eqref{equ:InpFNO} for NA input as discussed above. 
All BVPs have been implemented and solved numerically 
using spectral methods \citep{willot2015fourier,khorrami2020development} 
and the software toolkit DAMASK \citep{roters2019damask}. 
Assume now that \(n_{\mathrm{dat}}\) such BVPs have been solved on 
\(\mathrm{U}\) for 
\begin{equation}
E_{a}(\mathbf{x}_{b}),\nu_{a}(\mathbf{x}_{b}),\bar{\mathbf{F}}_{\!a}
\,,\quad
a=1,\ldots,n_{\mathrm{dat}},\,b=1,\ldots,n_{\mathrm{res}}
\,,
\label{equ:InpBVP}
\end{equation} 
yielding the corresponding discrete values 
\begin{equation}
\mathbf{P}_{\!ab}^{\mathrm{dat}}
:=\mathbf{P}_{\!a}^{\mathrm{dat}}(\mathbf{x}_{b})
\,,\ \ 
a=1,\ldots,n_{\mathrm{dat}},\,b=1,\ldots,n_{\mathrm{res}}
\,,
\label{equ:DatBVP}
\end{equation}
for the stress field. 
Here, \(n_{\mathrm{dat}}\) is the number of data fields for which values 
have been obtained at a finite number \(n_{\mathrm{res}}\) of points 
in \(\mathrm{U}\) ("res" stands for "resolution"). 
The grain microstructure in \(\mathrm{U}\) is represented by the spatial 
distribution of material property values in \eqref{equ:InpBVP}. 
Since these are constant within each grain 
(see e.g. Figure \ref{fig:ProMatDat} below), 
their spatial variation is determined by the grain morphology distribution 
in \(\mathrm{U}\) based on a corresponding mean grain size 
\(s_{\mathrm{U}}\) for the \(n_{\mathrm{dat}}\) data. Corresponding 
grain morphologies are generated via Voronoi tessellation. 

For simplicity, attention is restricted here to the special case of deformation 
in the \((x_{1},x_{2})\)-plane. In this case, 
\begin{equation}
\mathbf{F}
=\left\lbrack
\begin{array}{ccc}
F_{11}&F_{12}&0
\\
F_{21}&F_{22}&0
\\
0&0&1
\end{array}
\right\rbrack
\,,\quad
\mathbf{E}
=\left\lbrack
\begin{array}{ccc}
E_{11}&E_{12}&0
\\
E_{12}&E_{22}&0
\\
0&0&0
\end{array}
\right\rbrack
\,,\quad
\mathbf{P}
=\left\lbrack
\begin{array}{ccc}
P_{11}&P_{12}&0
\\
P_{21}&P_{22}&0
\\
0&0&P_{33}
\end{array}
\right\rbrack
\,,
\label{equ:DefPlaComCar}
\end{equation}
hold identically, the latter two in the context of 
\eqref{equ:ElaKirVenSai}-\eqref{equ:StnGreMatComCar}. Given 
\eqref{equ:DefPlaComCar}${}_{1,3}$, the reduced forms 
\begin{equation}
\begin{array}{rcl}
\mathbf{f}
&=&
(F_{\!11},F_{\!12},0,F_{\!21},F_{\!22},0,0,0,1)
\,,\\
\mathbf{p}
&=&
(P_{\!11},P_{\!12},0,P_{\!21},P_{\!22},0,0,0,P_{\!33})
\,,\\
\mathbf{a}
&=&
(\bar{A}_{11},\bar{A}_{12},\tilde{A}_{13},
\bar{A}_{21},\bar{A}_{22},\tilde{A}_{23},
\tilde{A}_{31},\tilde{A}_{32},\bar{A}_{33})
\,,\\
\imath\mathbf{k}\times\hat{\mathbf{a}}
&=&
\imath(k_{2}\hat{A}_{13},
-k_{1}\hat{A}_{13},0,
k_{2}\hat{A}_{23},
-k_{1}\hat{A}_{23},0,
0,0,k_{1}\hat{A}_{32}-k_{2}\hat{A}_{31})
\,,
\end{array}
\label{equ:AryDefPlaPcNAs}
\end{equation}
hold for the PcNA arrays introduced above.
The first of these for example determines \(\bar{\mathbf{f}}\) in the PcNA 
input \eqref{equ:InpFNO}; likewise, 
\begin{equation}
\textstyle
\mathbf{s}
=\left\lbrace
\begin{array}{lcl}
(P_{\!11},P_{\!12},0,P_{\!21},P_{\!22},0,0,0,P_{\!33})
&&
\hbox{PgNA, PiNA}
\\
(\bar{A}_{11},\bar{A}_{12},\tilde{A}_{13},
\bar{A}_{21},\bar{A}_{22},\tilde{A}_{23},
\tilde{A}_{31},\tilde{A}_{32},\bar{A}_{33})
&&
\hbox{PeNA}
\end{array}
\right.
\end{equation} 
for \(\mathbf{s}^{\mathrm{out}}\) in 
\eqref{equ:TraOutAuxPcFNO}-\eqref{equ:HidOutcFNO} as part of 
output transformation \(\bm{\tau}_{\mathrm{out}}\) is determined by 
\eqref{equ:AryDefPlaPcNAs}${}_{2,3}$.

For deformation in the \((x_{1},x_{2})\)-plane, 
numerical solution of the BVPs is based in particular on the 
uniform discretizations 
\begin{equation}
\begin{array}{rcl}
\mathrm{U}_{\mathrm{dis}}
&:=&
\lbrace
\mathbf{x}=(x_{1},x_{2},0)
\ \vert\ x_{d}=(i_{d}-1)\ell_{\mathrm{U}}/n_{\mathrm{dis}},
\,i_{d}=1,\ldots,n_{\mathrm{dis}},
\,d=1,2
\rbrace
\,,\\
\mathrm{U}_{\mathrm{dis}}^{\ast}
&:=&
\lbrace
\mathbf{k}=(k_{1},k_{2},0)
\ \vert\ k_{d}=2\pi(\mu_{d}-1)/\ell_{\mathrm{U}}
-\pi n_{\mathrm{dis}}/\ell_{\mathrm{U}},
\,\mu_{d}=1,\ldots,n_{\mathrm{dis}},
\,d=1,2
\rbrace 
\,,
\end{array}
\label{equ:CelUniDisFou}
\end{equation}
of \(\mathrm{U}\) and its reciprocal space \(\mathrm{U}^{\ast}\), respectively, 
for \(n_{\mathrm{dis}}\) even 
(recall \(n_{\mathrm{dis}}\leqslant n_{\mathrm{res}}\)), 
where \(\ell_{\mathrm{U}}\) is a typical lengthscale of \(\mathrm{U}\) 
(e.g., side length). Note that 
\(\mathbf{k}\cdot\mathbf{x}
=k_{d}x_{d}
=2\pi(i_{d}-1)(\mu_{d}-1)/n_{\mathrm{dis}}-\pi(i_{d}-1)\) 
in this case (summation convention). 
The discrete Fourier transform (DFT) corresponding to 
\eqref{equ:CelUniDisFou} is given by 
\begin{equation}
\textstyle
\check{f}(\mathbf{k})
:=n_{\mathrm{dis}}^{-2}
\sum_{\mathbf{x}\in\mathrm{U}_{\mathrm{dis}}}
e^{-\imath\mathbf{k}\cdot\mathbf{x}}
f(\mathbf{x})
\,,\quad
f(\mathbf{x})
=\check{f}(\bm{0})
+\sum_{\mathbf{k}\in\mathrm{U}_{\mathrm{dis}}^{\ast}
\setminus\lbrace\mathbf{0}\rbrace}
e^{\imath\mathbf{k}\cdot\mathbf{x}}
\check{f}(\mathbf{k})
\,,
\label{equ:TraForDisNA}
\end{equation} 
via trapezoidal approximation of \eqref{equ:FieScaSerFor}${}_{2}$ and 
truncation of \eqref{equ:FieScaSerFor}${}_{1}$ as usual.
This DFT is also employed in the FNO, and in particular to approximate the 
output transformations \eqref{equ:TraOutPcFNO}. For the current case of 
real-valued fields, the DFT represents the basis for the real-valued 
fast Fourier transform \cite[RFFT:~e.g.,][]{Sorensen1987} employed in the 
FNO \cite[e.g.,][]{Li2021}. 
For the uniform discretization \(n_{\mathrm{dis}}\) 
per dimension, the RFFT is based on \(\frac{1}{2}n_{\mathrm{dis}}+1\) 
independent modes per dimension. 

On this basis, \(n_{\mathrm{dat}}=1250\) 
stress field data with a resolution of \(n_{\mathrm{res}}=128\) 
have been obtained in \(\mathrm{U}\) subject to unixial extension in 
the \(x_{2}\) direction. In this case, 
\(\bar{F}_{\!11}=1\), 
\(\bar{F}_{\!22}\) is variable,
\(\bar{F}_{\!33}=1\), 
and \(\bar{F}_{\!i\!j}=0\) otherwise. Likewise, \(\bar{E}_{22}\) 
is then variable, and all other \(\bar{E}_{i\!j}=0\). 
Data are generated for \(E(\mathbf{x})\in\lbrack 50,200\rbrack\) GPa, 
\(\nu(\mathbf{x})\in\lbrack 0.25,0.35\rbrack\) (typical of metals), 
a mean grain size of \(s_{\mathrm{U}}=\frac{1}{3}\,\ell_{\mathrm{U}}\), 
\(\bar{F}_{\!22}=1.002\), and \(\bar{F}_{\!22}=1.004\). 
The latter for example corresponds to \(\bar{E}_{22}=0.4\%\) 
(all other \(\bar{E}_{i\!j}=0\)) for the mean Green strain. 
In the context of isotropic elasticity and random grain / material property 
distributions, \(P_{\!22}\) is then the largest stress component, 
as confirmed by the numerical results to follow. 

\subsection{NA parameter values, training and testing}

Values for a number of the NA and in particular FNO (hyper)parameters are 
adopted here for comparability from the previous work of \cite{Li2021} 
and \cite{Kovachki2023}, who considered (in the current context) only 
PgFNOs. To this end, note that their so-called "channel" dimensions 
\(d_{a}\), \(d_{v_{t}}\) (\(t=0,\ldots,T-1\)), and \(d_{u}\) 
represent the dimensions 
of the values taken by the input, hidden, and output, fields, 
respectively. Consequently, \(d_{a}\equiv d(\mathbf{i})\), 
\(d_{v_{t}}\equiv d(\mathbf{h}_{l})\), 
\(d_{u}\equiv d(\mathbf{o})\), and 
\(d_{v_{t}}/d_{a}\equiv n_{l}^{\mathrm{neu}}\) in the current notation. 
The numerical examples in \citet[][\S 5, Appendix]{Li2021} and 
\citet[][\S\S 6-7]{Kovachki2023} are based on scalar fields, in which case 
\(d_{a}=1=d_{u}\). They work with \(d_{v_{t}}=64\) in 1D, 
\(d_{v_{t}}=32\) in 2D, \(n_{\mathrm{hid}}=4\), and ReLU-based activation. 
For the current plane deformation (i.e., 2D) case, then, 
\(n_{l}^{\mathrm{neu}}=32\) and \(n_{\mathrm{hid}}=4\) are adopted 
here. Rather than on ReLU, the current 
\(\sigma\) is given by the GeLU function (i.e., a mollified ReLU function). 
Lastly, all computational examples to follow employ the non-activated NN form 
for \(\bm{\tau}_{\mathrm{inp}}\), i.e., 
\(\mathbf{h}_{0}
=\bm{\tau}_{\mathrm{inp}}\diamond\mathbf{i}
=\mathbf{W}_{\!\mathrm{inp}}\mathbf{i}\) 
(\(\mathbf{b}_{\mathrm{inp}}\equiv\mathbf{0}\)), 
as well as the NN form of 
\(\mathbf{s}^{\mathrm{out}}\) in \eqref{equ:HidOutcFNO} 
with \(\mathbf{b}_{\mathrm{out}}\equiv\mathbf{0}\) 
determining \(\bm{\tau}_{\mathrm{out}}\) in 
\eqref{equ:AppNeuLayMul}-\eqref{equ:OutHidInAppNeuLayMul}. 

Training and testing employ \(n_{\mathrm{tra}}\) 
and \(n_{\mathrm{tes}}\) stress field data, respectively, "sampled" 
at \(n_{\mathrm{dis}}\leqslant n_{\mathrm{res}}\) points in \(\mathrm{U}\) 
("dis" stands for "discretization") with 
\(n_{\mathrm{tra}}+n_{\mathrm{tes}}=n_{\mathrm{dat}}\) as usual. 
Data preparation for training and testing is based on min-max normalization of 
the data for \(E\) in \eqref{equ:InpBVP} and for \(\mathbf{P}^{\mathrm{dat}}\) 
in \eqref{equ:DatBVP}. 
Normalized (and so non-dimensional) values for 
\(\mathbf{P}^{\mathrm{dat}}\) are employed in the corresponding 
loss functions 
\begin{equation}
\textstyle
L(n):=\left\lbrace
\begin{array}{lcl}
L_{\mathrm{dat}}(n)
&&\hbox{Pg-, PeNA}
\\
L_{\mathrm{dat}}(n)+c_{\textrm{div}}L_{\mathrm{div}}(n)
&&\hbox{PiNA}
\end{array}
\right.
\label{equ:FunLosConPhy}
\end{equation} 
for training (\(n=n_{\mathrm{tra}}\)) and testing (\(n=n_{\mathrm{tes}}\)). 
The data part 
\begin{equation}
\textstyle
L_{\mathrm{dat}}(n)
:=\sqrt{
\dfrac{1}
{\sum_{a=1}^{n}\!\sum_{b=1}^{n_{\mathrm{dis}}}
|\mathbf{P}_{\!ab}^{\mathrm{dat}}|^{2}}
\sum_{a=1}^{n}\!\sum_{b=1}^{n_{\mathrm{dis}}}
|\mathbf{P}_{\!ab}^{\mathrm{out}}-\mathbf{P}_{\!ab}^{\mathrm{dat}}|^{2}
}
\label{equ:FunLosDat}
\end{equation} 
of \(L\) is based on relative \(L_{2}\) error with 
\(|\mathbf{P}_{\!ab}^{\mathrm{dat}}|^{2}
=\mathbf{P}_{\!ab}^{\mathrm{dat}}\cdot\mathbf{P}_{\!ab}^{\mathrm{dat}}\) and so on. 
On the other hand, 
the (mollified) absolute \(L_{2}\) error measure 
\begin{equation}
\textstyle
L_{\mathrm{div}}(n)
:=\sqrt{\epsilon
+\sum_{a=1}^{n}\!\sum_{b=1}^{n_{\mathrm{dis}}}
|\ell_{\mathrm{U}}\mathbf{d}_{ab}^{\smash{\mathrm{out}}}|^{2}
}
\label{equ:FunLosDiv}
\end{equation} 
(dimensionless) 
is chosen for the divergence-free constraint in the PiNA form 
\eqref{equ:FunLosConPhy}${}_{2}$ of \(L\) 
(recall \(\mathbf{d}=\mathop{\mathrm{div}}\mathbf{P}\) 
from \eqref{equ:StsDivRepFouAryCar}) with \(\epsilon\ll 1\) and 
\(|\ell_{\mathrm{U}}\mathbf{d}_{ab}^{\smash{\mathrm{out}}}|^{2}
=\ell_{\mathrm{U}}\mathbf{d}_{ab}^{\smash{\mathrm{out}}}
\cdot
\ell_{\mathrm{U}}\mathbf{d}_{ab}^{\smash{\mathrm{out}}}\). 
This constraint is weighted in \(L\) by the constraint coefficient 
\(c_{\textrm{div}}\) (dimensionless). The  
value of \(c_{\textrm{div}}\) determines the importance of 
\(L_{\mathrm{div}}\) relative to \(L_{\mathrm{dat}}\) in 
\eqref{equ:FunLosConPhy}${}_{2}$ for training and testing of 
the PiNA. In the context of \eqref{equ:FunLosConPhy}${}_{2}$, 
an increase in \(c_{\textrm{div}}\) will result in decreased 
\(|\mathbf{d}^{\mathrm{out}}|\) and increased  
\(|\mathbf{P}^{\mathrm{out}}-\mathbf{P}^{\mathrm{dat}}|\) 
after training and testing (i.e., minimization of \(L\)). This trade-off 
is documented in the computational examples below. 

Besides the choice of error measure, hyperparameters in 
\eqref{equ:FunLosConPhy}-\eqref{equ:FunLosDiv} include
(i) \(n_{\mathrm{tra}}\), 
(ii) \(n_{\mathrm{tes}}\), 
(iii) \(n_{\mathrm{dis}}\), and 
(iv) \(c_{\textrm{div}}\). 
Training and testing of the PcFNOs is based on \(n_{\mathrm{tra}}=1000\), 
\(n_{\mathrm{tes}}=250\), \(n_{\mathrm{dis}}=64\), 
ADAM optimization, 500 epochs for training, and an initial learning rate 
\(10^{-3}\) which is halved every 100 epochs. 

\subsection{Results}

\begin{figure}[H]
\centering
\begin{subfigure}[t]{0.3\textwidth}
\includegraphics[width=\textwidth]{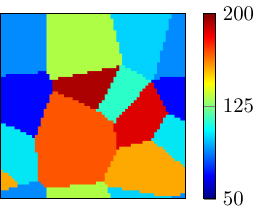}
\caption{\hbox{\(E(\mathbf{x})\) [GPa]}}
\end{subfigure}
\hspace{12mm}
\begin{subfigure}[t]{0.3\textwidth}
\includegraphics[width=\textwidth]{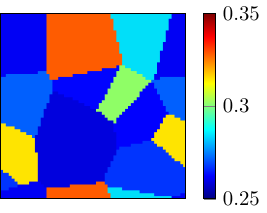}
\caption{\hbox{\(\nu(\mathbf{x})\)}}
\end{subfigure}
\caption{Grain microstructure in \(\mathrm{U}\) 
(\(x_{1}\) horizontal, \(x_{2}\) vertical, \(x_{3}=0\)) with 
\(E\in\lbrack 50,200\rbrack\) GPa, \(\nu\in\lbrack 0.25,0.35\rbrack\), 
\(n_{\mathrm{dis}}=64\), and 
\(s_{\mathrm{U}}=\frac{1}{3}\,\ell_{\mathrm{U}}\).
As discussed in the text, 
spatially varying material properties mimic an orientation 
microstructure. 
} 
\label{fig:ProMatDat}
\end{figure}

\begin{figure}[H]
\centering
\begin{subfigure}[t]{0.3\textwidth}
\includegraphics[width=\textwidth]{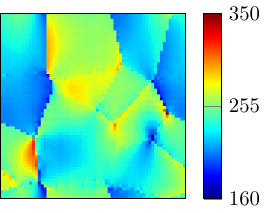}
\caption{\hbox{\(P_{\!11}^{\mathrm{dat}}\)}}
\vspace{3mm}
\end{subfigure}
\hspace{12mm}
\begin{subfigure}[t]{0.32\textwidth}
\includegraphics[width=\textwidth]{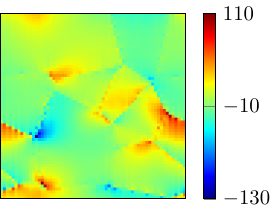}
\caption{\hbox{\(P_{\!12}^{\mathrm{dat}}\)}}
\vspace{3mm}
\end{subfigure}
\begin{subfigure}[t]{0.31\textwidth}
\includegraphics[width=\textwidth]{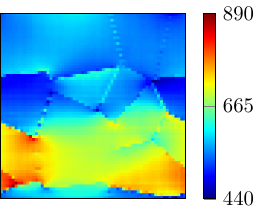}
\caption{\hbox{\(P_{\!22}^{\mathrm{dat}}\)}}
\end{subfigure}
\hspace{12mm}
\begin{subfigure}[t]{0.3\textwidth}
\includegraphics[width=\textwidth]{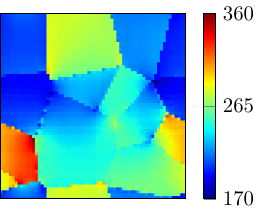}
\caption{\hbox{\(P_{\!33}^{\mathrm{dat}}\)}}
\end{subfigure}
%
\caption{Data for selected stress field components due to uniaxial extension 
of the polycrystal in Figure \ref{fig:ProMatDat} in the \(x_{2}\) (vertical) 
direction to \(\bar{F}_{\!22}=1.004\). 
Note the different scaling in each figure. 
Values in MPa. 
} 
\label{fig:DatStsComCar}
\end{figure}

\begin{figure}[H]
\centering
\begin{subfigure}[t]{0.3\textwidth}
\includegraphics[width=\textwidth]{output_channel_P22.pdf}
\caption{\hbox{data}}
\vspace{3mm}
\end{subfigure}
\hspace{12mm}
\begin{subfigure}[t]{0.3\textwidth}
\includegraphics[width=\textwidth]{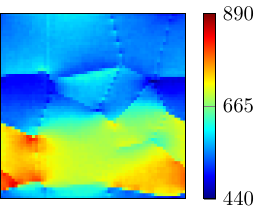}
\caption{\hbox{PgFNO}}
\vspace{3mm}
\end{subfigure}
\begin{subfigure}[t]{0.3\textwidth}
\includegraphics[width=\textwidth]{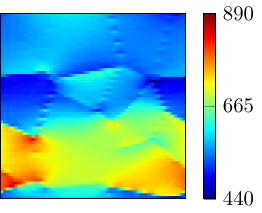}
\caption{\hbox{PiFNO, \(c_{\mathrm{div}}=0.01\)}}
\end{subfigure}
\hspace{12mm}
\begin{subfigure}[t]{0.3\textwidth}
\includegraphics[width=\textwidth]{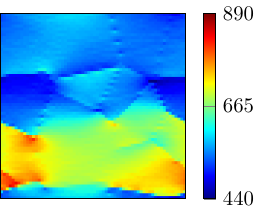}
\caption{\hbox{PeFNO}}
\end{subfigure}
\caption{Comparison of \(P_{\!22}^{\mathrm{dat}}\) (upper left) and 
\(P_{\!22}^{\mathrm{out}}\) from the PcFNOs. 
Values in MPa. 
}
\label{fig:P22_comparison}
\end{figure}

Comparison of results for the stress and stress divergence fields from the 
trained and tested PcFNOs (hereafter referred to simply as PcFNOs) 
with corresponding data is based on a grain microstructure which is not 
part of the training and testing data. 
Such a microstructure is shown in Figure \ref{fig:ProMatDat}. 
As shown, grain interiors are characterized by constant material properties whose values change discontinuously 
across grain boundaries. Corresponding stress field 
component data for uniaxial extension of the polycrystal in the vertical 
direction are displayed in Figure \ref{fig:DatStsComCar}. 
As discussed above, and evident 
in Figure \ref{fig:DatStsComCar}, \(P_{\!22}^{\mathrm{dat}}\) is the largest 
stress component in the data for uniaxial extension in the \(x_{2}\) direction. 
Analogous results for \(P_{\!22}^{\mathrm{out}}\) from the PcFNOs are 
compared with \(P_{\!22}^{\mathrm{dat}}\) in Figure \ref{fig:P22_comparison}.  
As evident from a comparison of Figures \ref{fig:ProMatDat}-\ref{fig:P22_comparison}, 
the stress field is relatively uniform in regions of uniform material properties 
(e.g., grain interiors), 
and varies in regions in which material properties change 
in a discontinuous fashion 
(e.g., across and around grain boundaries and triple junctions),  
resulting in stress field gradients. 

\begin{figure}[H]
\centering
\begin{subfigure}[t]{0.3\textwidth}
\includegraphics[width=\textwidth]{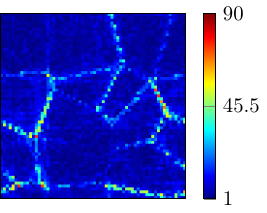}
\caption{\hbox{PgFNO}}
\vspace{3mm}
\end{subfigure}
\hspace{12mm}
\begin{subfigure}[t]{0.3\textwidth}
\includegraphics[width=\textwidth]{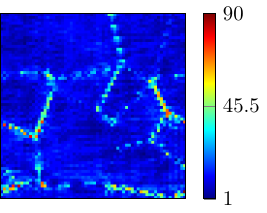}
\caption{\hbox{PeFNO}}
\vspace{3mm}
\end{subfigure}
\begin{subfigure}[t]{0.3\textwidth}
\includegraphics[width=\textwidth]{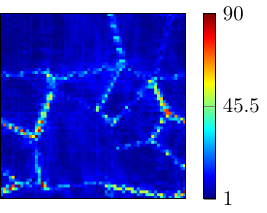}
\caption{\hbox{PiFNO, \(c_{\mathrm{div}}=0.01\)}}
\end{subfigure}
\hspace{12mm}
\begin{subfigure}[t]{0.3\textwidth}
\includegraphics[width=\textwidth]{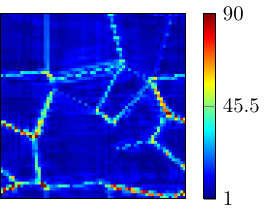}
\caption{\hbox{PiFNO, \(c_{\mathrm{div}}=0.1\)}}
\end{subfigure}
\caption{Magnitude 
\(|\mathbf{P}^{\mathrm{out}}-\mathbf{P}^{\mathrm{dat}}|\) 
of the error in \(\mathbf{P}^{\mathrm{out}}\). 
Values in MPa.
}
\label{fig:P22_dataError_comparison}
\end{figure}

\begin{figure}[H]
\centering
\begin{subfigure}[t]{0.35\textwidth}
\includegraphics[width=\textwidth]{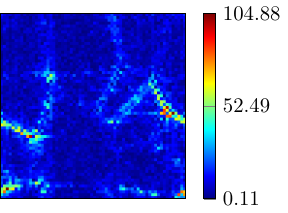}
\caption{\hbox{PgFNO}}
\vspace{3mm}
\end{subfigure}
\hspace{12mm}
\begin{subfigure}[t]{0.32\textwidth}
\includegraphics[width=\textwidth]{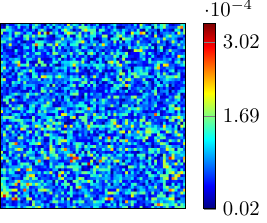}
\caption{\hbox{PeFNO}}
\vspace{3mm}
\end{subfigure}
\begin{subfigure}[t]{0.35\textwidth}
\includegraphics[width=\textwidth]{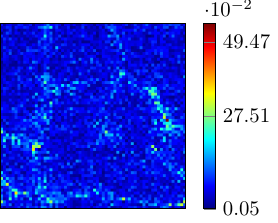}
\caption{\hbox{PiFNO, \(c_{\mathrm{div}}=0.01\)}}
\end{subfigure}
\hspace{10mm}
\begin{subfigure}[t]{0.35\textwidth}
\includegraphics[width=\textwidth]{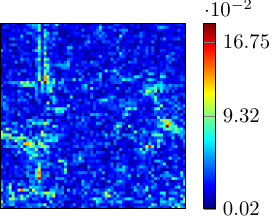}
\caption{\hbox{PiFNO, \(c_{\mathrm{div}}=0.1\)}}
\end{subfigure}
\caption{PcFNO results for the error \(|\mathbf{d}^{\mathrm{out}}|\) 
in \(\mathbf{d}^{\mathrm{out}}
=\mathop{\mathrm{div}}\mathbf{P}^{\mathrm{out}}\). 
Values in MPa/\(\ell_{\mathrm{U}}\). 
Note the different scaling in each figure.
}
\label{fig:ErrDivPcFNO}
\end{figure}

Since the data \eqref{equ:DatBVP} are uniformly distributed in \(\mathrm{U}\), 
note that there is less data in regions around grain boundaries and triple junctions 
where the stress field varies than in grain interiors where the stress field is relatively 
uniform. These properties of the stress field and data have an influence on, 
and are related to, the accuracy of 
\(\mathbf{P}^{\mathrm{out}}\) and 
\(\mathbf{d}^{\mathrm{out}}
=\mathop{\mathrm{div}}\mathbf{P}^{\mathrm{out}}\) from the PcFNOs. 
Recall that the latter is effectively calculated from the former in the context of 
\eqref{equ:TraOutAuxPcFNO}-\eqref{equ:TraDivOutPcFNO}. 
As a measure of the error in \(\mathbf{P}^{\mathrm{out}}\), 
consider the results for 
\(|\mathbf{P}^{\mathrm{out}}-\mathbf{P}^{\mathrm{dat}}|\) 
from the PcFNOs displayed in Figure \ref{fig:P22_dataError_comparison}. 
As evident, the errors in \(\mathbf{P}^{\mathrm{out}}\) 
from the PgFNO, PeFNO and PiFNO for \(c_{\mathrm{div}}\leqslant 0.01\) 
are quite small in the grain interiors (\(\approx 1\) MPa) and largest at 
grain boundaries and triple junctions. 
The errors in \(\mathbf{d}^{\mathrm{out}}\) 
from the PgFNO and PiFNO for \(c_{\mathrm{div}}\leqslant 0.1\) 
as measured by \(|\mathbf{d}^{\mathrm{out}}|\) are analogously 
distributed, as shown in Figures \ref{fig:ErrDivPcFNO}(a,c,d).
In contrast, \(|\mathbf{d}^{\mathrm{out}}|\) from the PeFNO 
(Figure \ref{fig:ErrDivPcFNO}(b)) is much more uniformly distributed 
in \(\mathrm{U}\). More significantly, the largest error in 
\(\mathbf{d}^{\mathrm{out}}\) from the PeFNO 
(Figure \ref{fig:ErrDivPcFNO}(b)) is three orders-of-magnitude 
smaller than the corresponding error in \(\mathbf{d}^{\mathrm{out}}\) 
from the PiFNO for \(c_{\mathrm{div}}\leqslant 0.1\) 
(Figures \ref{fig:ErrDivPcFNO}(c,d)) in the current example. 
Indeed, this is to be expected in the context of the output 
transformation \eqref{equ:TraOutPcFNO} with \eqref{equ:TraOutAuxPcFNO} 
and \eqref{equ:HidOutcFNO} of the current PeNAs, and from 
\eqref{equ:TraDivOutPcFNO} for \(\mathbf{d}^{\mathrm{out}}\). 

\begin{figure}[H]
\centering
\begin{subfigure}[t]{0.3\textwidth}
\includegraphics[width=\textwidth]{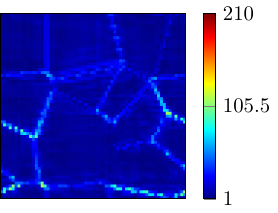}
\caption{\(|\mathbf{P}^{\mathrm{out}}-\mathbf{P}^{\mathrm{dat}}|\),
\(c_{\mathrm{div}}=0.1\)}
\vspace{3mm}
\end{subfigure}
\hspace{3mm}
\begin{subfigure}[t]{0.3\textwidth}
\includegraphics[width=\textwidth]{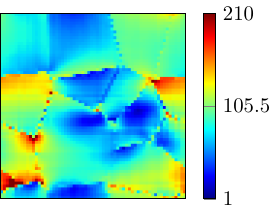}
\caption{\(|\mathbf{P}^{\mathrm{out}}-\mathbf{P}^{\mathrm{dat}}|\),
\(c_{\mathrm{div}}=1\)}
\vspace{3mm}
\end{subfigure}
\hspace{3mm}
\begin{subfigure}[t]{0.3\textwidth}
\includegraphics[width=\textwidth]{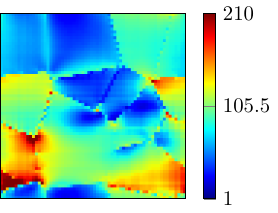}
\caption{\(|\mathbf{P}^{\mathrm{out}}-\mathbf{P}^{\mathrm{dat}}|\),
\(c_{\mathrm{div}}=10\)}
\vspace{3mm}
\end{subfigure}
\begin{subfigure}[t]{0.32\textwidth}
\includegraphics[width=\textwidth]{div_norm_3.pdf}
\caption{\(|\mathbf{d}^{\mathrm{out}}|\),
\(c_{\mathrm{div}}=0.1\)}
\end{subfigure}
\hspace{3mm}
\begin{subfigure}[t]{0.3\textwidth}
\includegraphics[width=\textwidth]{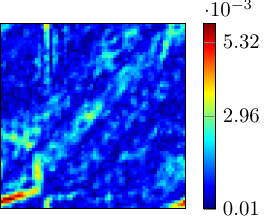}
\caption{\(|\mathbf{d}^{\mathrm{out}}|\),
\(c_{\mathrm{div}}=1\)}
\end{subfigure}
\hspace{3mm}
\begin{subfigure}[t]{0.3\textwidth}
\includegraphics[width=\textwidth]{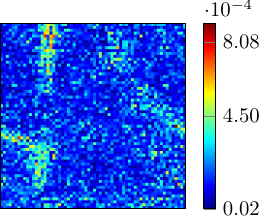}
\caption{\(|\mathbf{d}^{\mathrm{out}}|\),
\(c_{\mathrm{div}}=10\)}
\end{subfigure}
\caption{Comparison of PiFNO results for 
\(|\mathbf{P}^{\mathrm{out}}-\mathbf{P}^{\mathrm{dat}}|\) (above; units MPa)  
and \(|\mathbf{d}^{\mathrm{out}}|\) (below; units MPa/\(\ell_{\mathrm{U}}\)) 
for different values of \(c_{\mathrm{div}}\). 
Note the different scaling in each figure for \(|\mathbf{d}^{\mathrm{out}}|\). 
See text for discussion.
}
\label{fig:DivStsAccOffTraPiFNO}
\end{figure}

Comparison of Figures \ref{fig:P22_dataError_comparison}(c-d) and 
Figures \ref{fig:ErrDivPcFNO}(c-d) for the PiFNO implies 
that an increase in \(c_{\mathrm{div}}\) from \(0.01\) to \(0.1\) 
reduces the error in \(\mathbf{d}^{\mathrm{out}}\) without 
significantly affecting the accuracy of \(\mathbf{P}^{\mathrm{out}}\). 
This is no longer the case for larger values of \(c_{\mathrm{div}}\) 
as shown by the results in Figure \ref{fig:DivStsAccOffTraPiFNO}. 
Clearly, an increase in the accuracy of \(\mathbf{d}^{\mathrm{out}}\) 
from the PiFNO to the level of accuracy for \(\mathbf{d}^{\mathrm{out}}\) 
obtained by the PeFNO is possible 
(compare Figures \ref{fig:ErrDivPcFNO}(b) and \ref{fig:DivStsAccOffTraPiFNO}(f)), 
but only at the expense of a loss in the accuracy of 
\(\mathbf{P}^{\mathrm{out}}\) (Figures \ref{fig:DivStsAccOffTraPiFNO}(b,c)). 
As shown in Figure \ref{fig:DivStsAccOffTraPiFNO}, to achieve this, one is 
forced to overweight \(L_{\mathrm{div}}\) relative to \(L_{\mathrm{dat}}\)
in \eqref{equ:FunLosConPhy}${}_{2}$ during training and testing of the PiFNO. 
To discuss this briefly, recall that 
\(\mathbf{P}_{\!ab}^{\mathrm{dat}}\) 
for \(a=1,\ldots,n_{\mathrm{dat}}\), \(b=1,\ldots,n_{\mathrm{dis}}\), 
from \eqref{equ:DatBVP} 
are min-max normalized in the loss function \eqref{equ:FunLosConPhy}, resulting 
in a corresponding normalization of \(\mathbf{P}_{\!ab}^{\mathrm{out}}\) and 
\(\mathbf{d}_{ab}^{\mathrm{out}}\). Assume in this context that
\(|\mathbf{P}_{\!ab}^{\mathrm{dat}}|\lesssim 1\), 
\(|\mathbf{P}_{\!ab}^{\mathrm{out}}
-\mathbf{P}_{\!ab}^{\mathrm{dat}}|\lesssim 10^{-1}\), 
\(|\mathbf{P}_{\!ab}^{\mathrm{out}}|\lesssim 1\), and 
\(|\ell_{\mathrm{U}}\mathbf{d}_{ab}^{\mathrm{out}}|\lesssim 10^{-2}\) 
via \eqref{equ:StsDivRepFouAryCar} and \eqref{equ:TraForDisNA}, 
for \(a=1,\ldots,n_{\mathrm{dat}}\), \(b=1,\ldots,n_{\mathrm{dis}}\). 
Then \(L_{\mathrm{dat}}(n)\lesssim 10^{-1}\) and 
\(L_{\mathrm{div}}(n)\lesssim 10^{-2}\sqrt{n n_{\mathrm{dis}}}\sim 1\) for 
\(n=n_{\mathrm{tra}}=1000\) and \(n_{\mathrm{dis}}=64\). Consequently, 
\(L_{\mathrm{dat}}(n)\sim c_{\mathrm{div}}L_{\mathrm{div}}(n)\) for 
\(c_{\mathrm{div}}\sim 10^{-1}\), corresponding to the results in 
Figures \ref{fig:P22_dataError_comparison}(d) and \ref{fig:ErrDivPcFNO}(d). 
For \(c_{\mathrm{div}}\gtrsim 1\), then, \(L_{\mathrm{div}}(n)\) 
is overweighted, as implied by Figure \ref{fig:DivStsAccOffTraPiFNO}.

\begin{figure}[H]
\centering
\begin{subfigure}[t]{0.3\textwidth}
\includegraphics[width=\textwidth]{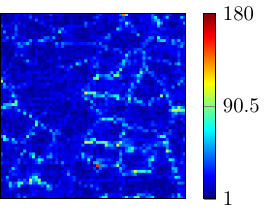}
\caption{\hbox{PgFNO}}
\end{subfigure}
\hspace{3mm}
\begin{subfigure}[t]{0.3\textwidth}
\includegraphics[width=\textwidth]{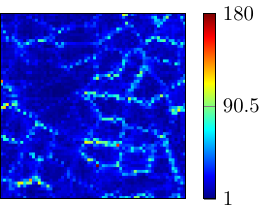}
\caption{\hbox{PeFNO}}
\end{subfigure}
\hspace{3mm}
\begin{subfigure}[t]{0.3\textwidth}
\includegraphics[width=\textwidth]{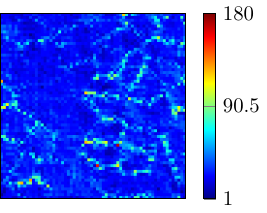}
\caption{\hbox{PiFNO}, \(c_{\mathrm{div}}=0.01\)}
\end{subfigure}
\caption{Magnitude 
\(|\mathbf{P}^{\mathrm{out}}-\mathbf{P}^{\mathrm{dat}}|\) of the error in 
\(\mathbf{P}^{\mathrm{out}}\) for a finer grain microstructure based on 
\(s_{\mathrm{U}}=\frac{1}{6}\,\ell_{\mathrm{U}}\) subject to uniaxial extension 
in the vertical direction with \(\bar{F}_{\!22}=1.004\).
Values in MPa. 
}
\label{fig:GraSizMeaExtRes}
\end{figure}

The results up to this point have been based on grain microstructures 
with a mean grain size of \(s_{\mathrm{U}}=\frac{1}{3}\,\ell_{\mathrm{U}}\) 
(e.g., Figure \ref{fig:ProMatDat}) deformed in uniaxial extension 
to \(\bar{F}_{\!22}=1.004\). 
As a last example, consider stress field results from the above PcFNOs 
(i.e., trained and tested with data based on 
\(s_{\mathrm{U}}=\frac{1}{3}\,\ell_{\mathrm{U}}\)) 
for a finer grain microstructure 
with mean grain size \(s_{\mathrm{U}}=\frac{1}{6}\,\ell_{\mathrm{U}}\), 
again undergoing uniaxial extension to \(\bar{F}_{\!22}=1.004\). Results 
for \(|\mathbf{P}^{\mathrm{out}}-\mathbf{P}^{\mathrm{dat}}|\) 
in this case are shown in Figure \ref{fig:GraSizMeaExtRes}. 
Similar to the results in Figure \ref{fig:P22_dataError_comparison}, the 
errors in the PgFNO, PeFNO, and PiFNO for \(c_{\mathrm{div}}=0.01\), 
are quite small in the grain interiors (\(\approx 1\) MPa) and largest at 
grain boundaries and triple junctions in Figure \ref{fig:GraSizMeaExtRes}. 
As discussed above in the context 
of the results in Figure \ref{fig:P22_dataError_comparison}, this is related 
to the fact that there is less data in regions where the stress field varies 
(e.g., grain boundaries and triple junctions) 
than in regions where the stress field is relatively uniform (e.g., grain interiors). 
Consequently, PcFNO output 
underestimates spatial variations in the stress field, especially near grain 
boundaries and triple junctions, resulting in the corresponding increase in 
error of the PcFNO ouput shown in Figure \ref{fig:P22_dataError_comparison} 
and Figure \ref{fig:GraSizMeaExtRes}. 


\section{Summary and outlook} 
\label{sec:OutSum}

A physics-encoded neural approximation (PeNA) has been developed 
in the current work for the data-based modeling of quasi-static equilibrium 
stress fields in solids. For the corresponding constraint of divergence-free stress, 
a novel encoding approach based on a stress potential is proposed. As shown 
in the current work, inclusion ("encoding") of this constraint in the NA 
architecture rather than in the loss function as done in the physics-informed 
case yields more accurate and robust NA output for the equillibrium stress 
field. This is also related to the fact that only the training and testing of the 
PiNA is constrained by mechanical equilibrium; in contrast, it constrains both 
the training and output of the PeNA. 

One means of reducing error in the PcFNO output as shown in 
Figure \ref{fig:P22_dataError_comparison} and Figure \ref{fig:GraSizMeaExtRes} 
would be of course to increase the number and density of data points 
at and around grain boundaries and triple junctions. Another would 
be to train and test the FNOs with data on the stress field and its gradient field.
A third possibility to this end 
would be non-uniform weighting of the data during training and testing. 
As discussed above, training and testing of the PcFNO based 
on \eqref{equ:FunLosDat} and \eqref{equ:FunLosDiv} tacitly assumes 
uniform weighting. This can be seen by expressing these in weighted form 
\begin{equation}
\begin{array}{rcl}
L_{\mathrm{dat}}(n)
&=&
\sqrt{
\sum_{a=1}^{n}\!\sum_{b=1}^{n_{\mathrm{dis}}}
(\mathbf{P}_{\!ab}^{\mathrm{out}}-\mathbf{P}_{\!ab}^{\mathrm{dat}})
\cdot
\mathbf{W}_{\!ab}^{\mathrm{dat}}
\,(\mathbf{P}_{\!ab}^{\mathrm{out}}-\mathbf{P}_{\!ab}^{\mathrm{dat}})}
\,,\\
c_{\textrm{div}}L_{\mathrm{div}}(n)
&=&
\sqrt{
\sum_{a=1}^{n}\!\sum_{b=1}^{n_{\mathrm{dis}}}
\ell_{\mathrm{U}}\mathbf{d}_{ab}^{\smash{\mathrm{out}}}
\cdot 
\mathbf{W}_{\!ab}^{\mathrm{div}}
\,\ell_{\mathrm{U}}\mathbf{d}_{ab}^{\smash{\mathrm{out}}}}
\,,
\end{array}
\label{equ:FunLosWei}
\end{equation} 
where 
\begin{equation}
\begin{array}{rclcrcl}
\mathbf{W}_{\!ab}^{\mathrm{dat}}
&=&
w_{\mathrm{dat}}\,\mathbf{I}
\,,&&
w_{\mathrm{dat}}
&=&
\dfrac{1}
{\sum_{a=1}^{n}\!\sum_{b=1}^{n_{\mathrm{dis}}}
|\mathbf{P}_{\!ab}^{\mathrm{dat}}|^{2}}
\,,\\
\mathbf{W}_{\!ab}^{\mathrm{div}}
&=&
w_{\mathrm{div}}\mathbf{I}
\,,&&
w_{\mathrm{div}}
&=&
c_{\textrm{div}}^{2}
\,,
\end{array}
\label{equ:FunLosWeiUni}
\end{equation} 
represent the corresponding weight matrices. To compensate for the lack 
of data in regions with large stress gradients, one could model each component 
of \(\mathbf{W}_{\!ab}\) proportional to the magnitude 
of the (non-dimensional) gradient of 
the corresponding component of \(\mathbf{P}_{\!a}^{\textrm{dat}}\) 
at \(\mathbf{x}_{b}\). As a result, data from regions with large stress gradients 
would have a greater influence on training and testing of the PcFNOs based 
on \eqref{equ:FunLosWei} than 
data from regions of uniform stress, resulting in a decrease of error at 
grain boundaries and triple junctions. 

The neural approximations considered in the current work can be improved 
and refined in a number of ways. 
One possibility here is extension of the 
NA input \(\mathbf{i}(\mathbf{x})\) in \eqref{equ:InpFNO} to include 
the deformation gradient field \(\mathbf{F}\), i.e., 
\(\mathbf{i}(\mathbf{x})
=(E(\mathbf{x}),\nu(\mathbf{x}),\mathbf{f}(\mathbf{x}))\). 
In this case, one could also include angular momentum balance 
\eqref{equ:BalMomQuaStaNonGeo}${}_{2}$ in the constraint for 
quasi-static mechanical equilibrium. 
Another possibility concerns the architecture of the PeNA for 
divergence-free stress based on the stress potential \(\mathbf{A}\). 
Since \(\tilde{\mathbf{P}}=\mathop{\mathrm{curl}}\tilde{\mathbf{A}}\) 
is not invertible, calculating \(\tilde{\mathbf{A}}\) as the 
"anti-curl" of \(\tilde{\mathbf{P}}\) is non-unique and a further 
source of error in operator training. 
To avoid this, further constraints on \(\tilde{\mathbf{A}}\) are necessary. 
The most common of these is the Coulomb gauge condition 
\(\mathop{\mathrm{div}}\tilde{\mathbf{A}}=\bm{0}\). 
In this case, \(\tilde{\mathbf{P}}=\mathop{\mathrm{curl}}\tilde{\mathbf{A}}\) 
is invertible. A corresponding improved PeNA is obtained by 
accounting for this condition as a constraint in training and testing. 

These and other possible further developments of PeNAs 
for data-based computational modeling in solid mechanical represent 
work in progress to be reported on in the future. 

\bibliographystyle{elsarticle-harv}

\bibliography{References}

\begin{appendix}



\section{Helmholtz \& Hodge decompositions} 
\label{app:DecHH}

The mathematical basis for the stress potential employed in the 
physics-encoded neural approximations for quasi-static mechanical equilibrium is 
the Helmholtz decomposition of vector fields \cite[e.g.,][]{Bhatia2013}. 
This is briefly summarized in the following and compared 
with the related Hodge decomposition of 1-form fields employed 
by \cite{RichterPowell2022} to encode divergence-free vector fields in 
NN architectures. 
Central to the current work is a generalization of the Helmholtz decomposition 
for vector fields to second-order tensor fields also treated in the following. 

For simplicity, attention is restricted here to fields on unbounded domains. 

\subsection*{Helmholtz decomposition of vector fields} 

Let \(\nabla\) represent the Euclidean gradient operator, and \(\bm{u}\) 
a differentiable vector field. The gradient \(\nabla\bm{u}\) of \(\bm{u}\) 
determines as usual the divergence and curl of \(\bm{u}\), defined by
\begin{equation}
\mathop{\mathrm{div}}\bm{u}
:=\bm{I}\cdot\nabla\bm{u}
\,,\quad
\bm{c}\cdot\mathop{\mathrm{curl}}\bm{u}
:=\mathop{\mathrm{div}}\bm{u}\times\bm{c}
=\mathop{\mathrm{div}}\,(\mathop{\mathrm{axt}}\bm{u})\,\bm{c}
\,,
\label{equ:FieVecCurDivCalEuc}
\end{equation} 
respectively, 
\cite[e.g.,][Chapter 1]{Chadwick1999} for any constant \(\bm{c}\) 
(unless restricted by parentheses, all operators apply to everything on their right).
Given these, the Helmholtz decomposition of \(\bm{u}\) takes the form 
\begin{equation}
\bm{u}=\nabla\varsigma+\mathop{\mathrm{curl}}\bm{\varphi}
\label{equ:FieVecDecHel}
\end{equation} 
\cite[e.g.,][]{Bhatia2013} 
determined by scalar \(\varsigma\) and vector \(\bm{\varphi}\) potentials. 
Properties of this split and the potentials include
\vspace{-3mm}
\begin{itemize}
\itemsep-1pt
\item \(\varsigma\) and \(\bm{\varphi}\) are determined only up to constants, 
\item \(\bm{\varphi}+\nabla f\) is also a vector potential for any smooth 
scalar field \(f\), 
\item \(\bm{u}\) is divergence-free for \(\varsigma\) harmonic, i.e., 
\(\mathop{\mathrm{div}}\nabla\varsigma=0\). 
\end{itemize} 
\vspace{-2mm}
In particular, the trivial scalar potential 
\begin{equation}
\varsigma=\mathrm{const.}
\label{equ:FieVecPotScaTri}
\end{equation}
determines the corresponding special case 
\begin{equation}
\bm{u}=\mathop{\mathrm{curl}}\bm{\varphi}
\label{equ:FieVecFreDiv}
\end{equation} 
of \eqref{equ:FieVecDecHel} relevant to the encoding of divergence-free 
vector fields in NN and NO architectures. 

\subsection*{Hodge decomposition of 1-form fields} 

To encode divergence-free vector fields in NN architectures, 
\cite{RichterPowell2022} work with 
the Hodge decomposition of 1-form fields \cite[e.g.,][]{Abr88} 
rather than with \eqref{equ:FieVecDecHel}. 
Although they consider the \(n\)-dimensional case, 
for comparison with the current approach, 
attention is restricted here for simplicity to the physically relevant case \(n=3\). 

Let \(g\) represent the standard Euclidean metric with 
\(g(\bm{a},\bm{b})=\bm{a}\cdot\bm{b}\), 
and \(\omega\) the standard 
Euclidean triple product with \(\omega(\bm{a},\bm{b},\bm{c})
=g(\bm{a}\times\bm{b},\bm{c})\). 
The interior product operator \(\imath_{\bm{a}}\) \cite[e.g.,][\S 5.1]{Abr88} 
maps \(g\) to the 1-covector \(\imath_{\bm{a}}g\) defined by 
\((\imath_{\bm{a}}g)(\bm{b}):=g(\bm{a},\bm{b})\), and \(\omega\) 
to the 2-covector \(\imath_{\bm{a}\,}\omega\) defined by 
\((\imath_{\bm{a}\,}\omega)(\bm{b},\bm{c}):=\omega(\bm{a},\bm{b},\bm{c})\). 
The Hodge star operator \(\star\) \cite[e.g.,][Definition 6.2.12]{Abr88} 
maps \(k\)-covectors (\(k=0,\ldots,3\)) to \(3-k\) covectors and 
\(\star\star=(-1)^{k(3-k)}\). In particular, note that \(\star 1=\omega\), 
\(\star\omega=1\), 
\(\star\imath_{\bm{c}}g=\imath_{\bm{c}\,}\omega\) and 
\(\star\imath_{\bm{c}\,}\omega=\imath_{\bm{c}}g\). 
In terms of these operations, the Hodge decomposition of the 1-form 
field \(\imath_{\bm{u}}g\) dual to \(\bm{u}\) takes the form 
\begin{equation}
\imath_{\bm{u}}g=d\varsigma+\star d\imath_{\bm{\varphi}}g
=\imath_{\nabla\varsigma}g+\imath_{\mathop{\mathrm{curl}}\bm{\varphi}}g
\,,
\label{equ:FieForDecHog}
\end{equation} 
where \(\varsigma\) and \(\bm{\varphi}\) are the same potentials as 
in \eqref{equ:FieVecDecHel}, and \(d\) is the exterior derivative operator 
\cite[e.g.,][Chapter 6]{Abr88}. To be precise, 
\cite{RichterPowell2022} work with the axial tensor 
\(\mathop{\mathrm{axt}}\bm{\varphi}\) of \(\bm{\varphi}\) 
(their \(\bm{A}\)) as well as the alternative vector potential \(\bm{b}\) with 
\(\star d\!\star\!\imath_{\bm{b}\,}\omega=\imath_{\bm{\varphi}}g\). 
Choice of the trivial scalar potential \eqref{equ:FieVecPotScaTri} reduces 
\eqref{equ:FieForDecHog} to 
\begin{equation}
\imath_{\bm{u}}g=\star d\imath_{\bm{\varphi}}g
\label{equ:FieForFreDiv}
\end{equation} 
which is identically divergence-free since then 
\(
\mathop{\mathrm{div}}\bm{u}
=\star d\imath_{\bm{u}\,}\omega
=\star dd\imath_{\bm{\varphi}}g
=0\). 

For vector fields \(\bm{u}\) and their 1-form duals 
\(\imath_{\bm{u}}g\), the Helmholtz \eqref{equ:FieVecDecHel} 
and Hodge \eqref{equ:FieForDecHog} decompositions are  
equivalent. In contrast to the Helmholtz case, however, the Hodge 
decomposition is not readily generalizable 
to higher- and in particular second-order tensor fields like the stress. 

\subsection*{Generalization of the Helmholtz decomposition to tensor fields}

This is obtained from the vector form \eqref{equ:FieVecDecHel} as follows. 
Any differentiable second-order tensor field \(\bm{S}\) 
and constant vector \(\bm{c}\) induce a corresponding vector field 
\(\bm{v}=\bm{S}^{\mathrm{T}}\!\bm{c}\) whose divergence 
and curl are defined by \eqref{equ:FieVecCurDivCalEuc}. Using these, 
one can define the corresponding operators 
\begin{equation}
\bm{c}\cdot\mathop{\mathrm{div}}\bm{S}
:=\mathop{\mathrm{div}}\bm{S}^{\mathrm{T}}\!\bm{c}
\,,\quad
(\mathop{\mathrm{curl}}\bm{S})^{\mathrm{T}}\bm{c}
:=\mathop{\mathrm{curl}}\bm{S}^{\mathrm{T}}\!\bm{c}
\,,
\label{equ:FieTenCurDivCalEuc}
\end{equation} 
on \(\bm{S}\). 
Analogously, substitution of 
\(\bm{u}=\bm{S}^{\mathrm{T}}\bm{c}\), 
\(\varsigma=\bm{\phi}\cdot\bm{c}\), and 
\(\bm{\varphi}=\bm{\Phi}^{\mathrm{T}}\bm{c}\) 
into the vector Helmholtz decomposition \eqref{equ:FieVecDecHel} 
yields its generalization to second-order tensor fields 
\begin{equation}
\bm{S}=\nabla\bm{\phi}+\mathop{\mathrm{curl}}\bm{\Phi}
\label{equ:FieTenHelSto}
\end{equation} 
via \eqref{equ:FieTenCurDivCalEuc}${}_{2}$ and the identity 
\(\nabla(\bm{\phi}\cdot\bm{c})=(\nabla\bm{\phi})^{\mathrm{T}}\bm{c}\), 
again for constant \(\bm{c}\). The PeFNO in the paper 
for divergence-free stress employs the special case 
\begin{equation}
\bm{\phi}=\mathrm{const.}
\quad\Longrightarrow\quad
\bm{S}=\mathop{\mathrm{curl}}\bm{\Phi}
\label{equ:FieTenGenFreDiv}
\end{equation} 
of \eqref{equ:FieTenHelSto} analogous to \eqref{equ:FieVecFreDiv}. 


\section{Divergence-free stress based on fourth-order tensor fields}
\label{app:RTNN}


\subsection*{Symmetric stress fields based on Riemann tensors}

As discussed in the Introduction, \cite{Jnini2025} have recently introduced 
Riemann tensor neural networks (RTNNs) to encode divergence-free symmetric 
second-order tensor fields in a NN architecture. This is applied by them to 
mass and (linear) momentum balance in computational fluid dynamics and 
magnetohydrodynamics. For simplicity, the treatment here 
is limited to quasi-static linear momentum balance. In this context, 
the approach of \cite{Jnini2025} applies to symmetric stress tensor fields 
and the corresponding form \eqref{equ:BalMomQuaStaLinGeo} of 
quasi-static momentum balance relevant to (geometrically) linear solid mechanics. 

As the name implies, RTNNs are based on so-called Riemann tensors, i.e., 
fourth-order tensors \(\msbi{R}=R_{ijkl\,}
\bm{i}_{i}\otimes\bm{i}_{\!j}\otimes\bm{i}_{k}\otimes\bm{i}_{l}\) 
(summation convention) having the symmetry properties of the Riemann 
curvature tensor from differential geometry, i.e.,  
\begin{equation}
\msbi{R}^{\mathrm{T}}=\msbi{R}
\,,\quad
\msbi{R}\bm{A}=\msbi{R}\,\mathop{\mathrm{skw}}\bm{A}
\quad\Longrightarrow\quad
\msbi{R}^{\mathrm{T}}\!\bm{A}
=\msbi{R}^{\mathrm{T}}\mathop{\mathrm{skw}}\bm{A}
\label{equ:TenRieProSym}
\end{equation}
for all second-order tensor \(\bm{A}\). In these relations, 
\(\msbi{R}^{\mathrm{T}}\) represents the (major) transpose of \(\msbi{R}\). 
For any fourth-order tensor \(\msbi{A}\), this is defined by 
\begin{equation}
\msbi{A}^{\mathrm{T}}\bm{B}\cdot\bm{C}
:=\bm{B}\cdot\msbi{A}\bm{C}
\label{equ:MajTraTenOrdFou}
\end{equation} 
for all second-order tensors \(\bm{B},\bm{C}\). 

In this context, let \(\msbi{K}\) represent a Riemann tensor field, i.e., 
a fourth-order tensor field satisfying \eqref{equ:TenRieProSym}. Note that 
\(\msbi{K}\) determines a second-order tensor field \(\bm{T}\) via 
\begin{equation}
\bm{T}\bm{b}
:=\mathop{\mathrm{div}}\,(\mathop{\mathrm{div}}\msbi{K})\,\bm{b}
=(\nabla_{\!\bm{i}_{l}}\nabla_{\!\bm{i}_{m}}\msbi{K})
\,\lbrack\bm{b}\otimes\bm{i}_{m}\rbrack\,\bm{i}_{l}
\label{equ:TenRieSts}
\end{equation} 
(summation convention) for all constant \(\bm{b}\) with 
\(\nabla_{\!\bm{i}_{k}}\msbi{K}:=(\nabla\msbi{K})\,\bm{i}_{k}\). 
Equivalently, 
\begin{equation}
\bm{a}\cdot\bm{T}\bm{b}
=\bm{a}\otimes\bm{i}_{l}
\cdot
(\nabla_{\!\bm{i}_{l}}\nabla_{\!\bm{i}_{m}}\msbi{K})
\,\lbrack\bm{b}\otimes\bm{i}_{m}\rbrack
\label{equ:ScaTenRieSts}
\end{equation} 
for all constant \(\bm{a}\) and \(\bm{b}\). Since 
\(\msbi{K}^{\mathrm{T}}=\msbi{K}\) via \eqref{equ:TenRieProSym}${}_{1}$, 
\eqref{equ:ScaTenRieSts} implies \(\bm{T}^{\mathrm{T}}=\bm{T}\), i.e., 
\(\bm{T}\) symmetric. 
The divergence of \eqref{equ:TenRieSts} takes the form 
\begin{equation}
\bm{b}\cdot\mathop{\mathrm{div}}\bm{T}
=\bm{i}_{k}\vee\bm{i}_{l}
\cdot
(\nabla_{\!\bm{i}_{k}}\nabla_{\!\bm{i}_{l}}\nabla_{\!\bm{i}_{m}}\msbi{K})
\,\lbrack\bm{b}\otimes\bm{i}_{m}\rbrack
=\bm{b}
\cdot
(\nabla_{\!\bm{i}_{k}}\nabla_{\!\bm{i}_{l}}\nabla_{\!\bm{i}_{m}}\msbi{K})
\,\lbrack\mathop{\mathrm{skw}}\bm{i}_{l}\vee\bm{i}_{m}\rbrack\,\bm{i}_{k}
\label{equ:TenRieStsDiv}
\end{equation}
via \eqref{equ:FieTenCurDivCalEuc}${}_{1}$, 
the symmetry of \(\bm{T}\), the Euler symmetry 
\(\nabla_{\!\bm{i}_{k}}\nabla_{\!\bm{i}_{l}}
=\nabla_{\!\bm{i}_{l}}\nabla_{\!\bm{i}_{k}}\), 
and \eqref{equ:TenRieProSym}. Since 
\(\mathop{\mathrm{skw}}\bm{i}_{l}\vee\bm{i}_{m}=\bm{0}\) identically, 
and \(\bm{b}\) is arbitrary, \(\bm{T}\) as defined by \eqref{equ:TenRieSts} is 
both symmetric and divergence-free for \(\msbi{K}\) satisfying 
\eqref{equ:TenRieProSym}. 

Analogous to \eqref{equ:StsPotSerForFluMea} and 
\eqref{equ:StsKirPioFirPotSerForCur}${}_{1,2}$
in the text, the above relations can be expressed in Fourier series form 
\begin{equation}
\begin{array}{rclcl}
\msbi{K}(\bm{x})
&=&
\skew3\bar{\msbi{K}}+\skew3\tilde{\msbi{K}}(\bm{x})
&=&
\skew3\hat{\msbi{K}}(\bm{0})
+\sum_{\bm{k}\neq\bm{0}}
e_{}^{\imath{}\bm{k}\cdot\bm{x}}
\skew3\hat{\msbi{K}}(\bm{k})
\,,\\
\mathop{\mathrm{div}}\msbi{K}(\bm{x})
&=&
\mathop{\mathrm{div}}\skew3\tilde{\msbi{K}}(\bm{x})
&=&
\sum_{\bm{k}\neq\bm{0}}
e_{}^{\imath{}\bm{k}\cdot\bm{x}}
\skew3\hat{\msbi{K}}(\bm{k})
\,\imath\bm{k}
\,.
\end{array}
\end{equation} 
Given these, 
\begin{equation}
\begin{array}{rcl}
\bm{a}\cdot\bm{T}(\bm{x})\,\bm{b}
&=&
\sum_{\bm{k}\neq\bm{0}}
e_{}^{\imath{}\bm{k}\cdot\bm{x}}
\bm{a}\otimes\imath\bm{k}
\cdot
\skew3\hat{\msbi{K}}(\bm{k})\,\lbrack\bm{b}\otimes\imath\bm{k}\rbrack
\,,\\
\mathop{\mathrm{div}}\bm{T}(\bm{x})
&=&
\sum_{\bm{k}\neq\bm{0}}
e_{}^{\imath{}\bm{k}\cdot\bm{x}}
\skew3\hat{\msbi{K}}(\bm{k})
\,\lbrack\mathop{\mathrm{skw}}\imath\bm{k}\otimes\imath\bm{k}\rbrack
\,\imath\bm{k}
\,,
\end{array}
\label{equ:DivStsTenRieSerFor}
\end{equation} 
then follow from \eqref{equ:ScaTenRieSts} and 
\eqref{equ:TenRieStsDiv} via \eqref{equ:TenRieProSym}. In particular, since 
\(\mathop{\mathrm{skw}}\imath\bm{k}\otimes\imath\bm{k}=\bm{0}\) 
identically, \(\bm{T}\) given by \eqref{equ:DivStsTenRieSerFor}${}_{1}$ 
is both symmetric and divergence-free. 

Note that any fourth-order tensor with right and left minor 
or skew-symmetries like \(\msbi{K}\) has only 9 independent 
components, i.e., 
\(K_{ijkl}=\bm{i}_{i}\wedge\bm{i}_{\!j}
\cdot
\msbi{K}\lbrack\bm{i}_{k}\wedge\bm{i}_{l}\rbrack\). The additional 
major symmetry \eqref{equ:TenRieProSym}${}_{1}$ of \(\msbi{K}\) 
reduces these to 6 independent components. This corresponds to the 
number of independent components of \(\bm{T}\). 


\subsection*{Generalization to non-symmetric stress fields}

Although not considered by \cite{Jnini2025}, it turns out that one can 
generalize their approach to non-symmetric linear momentum flux 
densities, i.e., to the first Piola-Kirchhoff stress \(\bm{P}\) in 
\eqref{equ:BalMomQuaStaNonGeo}. Indeed, one need only relax 
the major symmetry assumption \eqref{equ:TenRieProSym}${}_{1}$. 
To show this, let \(\msbi{S}\) be a fourth-order Euclidean tensor field and 
\(\mathop{\mathrm{div}}\msbi{S}=(\nabla_{\!\bm{i}_{i}}\msbi{S})\bm{i}_{i}\) 
(summation convention) as above. Then 
\begin{equation}
\begin{array}{rclcl}
(\mathop{\mathrm{div}}\msbi{S})\,\bm{a}
&=&
((\nabla_{\!\bm{i}_{m}}\msbi{S})\bm{i}_{m})\bm{a}
&=&
(\nabla_{\!\bm{i}_{m}}\msbi{S})
\,\lbrack\bm{a}\otimes\bm{i}_{m}\rbrack
\,,\\
\mathop{\mathrm{div}}\,(\mathop{\mathrm{div}}\msbi{S})\,\bm{a}
&=&
(((\nabla_{\!\bm{i}_{l}}\nabla_{\!\bm{i}_{m}}\msbi{S})
\bm{i}_{m})\bm{a})\bm{i}_{l}
&=&
(\nabla_{\!\bm{i}_{l}}\nabla_{\!\bm{i}_{m}}\msbi{S})
\,\lbrack\bm{a}\otimes\bm{i}_{m}\rbrack\bm{i}_{l}
\,,
\end{array}
\label{equ:RelDivTenRie}
\end{equation}
for all constant \(\bm{a}\). Defining the first Piola-Kirchhoff stress 
\(\bm{P}\) in \eqref{equ:BalMomQuaStaNonGeo} by 
\begin{equation}
\bm{P}^{\mathrm{T}}\!\bm{a}
:=\mathop{\mathrm{div}}\,(\mathop{\mathrm{div}}\msbi{S})\,\bm{a}
\,,
\label{equ:FieTenStsRie}
\end{equation}
one then obtains  
\begin{equation}
\begin{array}{rclcl}
\bm{a}\cdot\bm{P}\bm{b}
&=&
\bm{b}\otimes\bm{i}_{l}
\cdot
(\nabla_{\!\bm{i}_{l}}\nabla_{\!\bm{i}_{m}}\msbi{S})
\,\lbrack\bm{a}\otimes\bm{i}_{m}\rbrack
&=&
\bm{a}\otimes\bm{i}_{l}
\cdot
(\nabla_{\!\bm{i}_{l}}\nabla_{\!\bm{i}_{m}}\msbi{S}^{\mathrm{T}})
\,\lbrack\bm{b}\otimes\bm{i}_{m}\rbrack
\,,\\
\bm{a}\cdot\mathop{\mathrm{div}}\bm{P}
&=&
\bm{i}_{k}\vee\bm{i}_{l}
\cdot
(\nabla_{\!\bm{i}_{k}}\nabla_{\!\bm{i}_{l}}\nabla_{\!\bm{i}_{m}}\msbi{S})
\,\lbrack\bm{a}\otimes\bm{i}_{m}\rbrack
&=&
\bm{a}
\cdot
(\nabla_{\!\bm{i}_{k}}\nabla_{\!\bm{i}_{l}}\nabla_{\!\bm{i}_{m}}
\msbi{S}^{\mathrm{T}})
\,\lbrack\bm{i}_{k}\vee\bm{i}_{l}\rbrack\,\bm{i}_{m}
\,,
\end{array}
\end{equation}
via \eqref{equ:FieTenCurDivCalEuc}${}_{1}$, 
\eqref{equ:MajTraTenOrdFou}, and the Euler symmetry 
\(\nabla_{\!\bm{i}_{l}}\nabla_{\!\bm{i}_{k}}
=\nabla_{\!\bm{i}_{k}}\nabla_{\!\bm{i}_{l}}\). Consequently, 
\begin{equation}
\mathop{\mathrm{div}}\bm{P}
=(\nabla_{\!\bm{i}_{k}}
\nabla_{\!\bm{i}_{l}}
\nabla_{\!\bm{i}_{m}}\msbi{S}^{\mathrm{T}})
\,\lbrack\bm{i}_{k}\vee\bm{i}_{l}\rbrack\,\bm{i}_{m}
\,.
\label{equ:FieTenStsDivGen}
\end{equation} 
Assume next that \(\msbi{S}\) is left-minor skew-symmetric or antisymmetric, i.e.,  
\begin{equation}
\msbi{S}^{\mathrm{T}}\!\bm{A}
=\msbi{S}^{\mathrm{T}}\!\mathop{\mathrm{skw}}\bm{A}
\label{equ:SymSkwLef}
\end{equation}
for all \(\bm{A}\). Then 
\begin{equation}
\mathop{\mathrm{div}}\bm{P}
=(\nabla_{\!\bm{i}_{k}}
\nabla_{\!\bm{i}_{l}}
\nabla_{\!\bm{i}_{m}}\msbi{S}^{\mathrm{T}})
\,\lbrack\mathop{\mathrm{skw}}\bm{i}_{k}\vee\bm{i}_{l}\rbrack\,\bm{i}_{m}
\label{equ:FieTenStsDiv}
\end{equation} 
follows from \eqref{equ:FieTenStsDivGen}. Since 
\(\mathop{\mathrm{skw}}\bm{i}_{k}\vee\bm{i}_{l}=\bm{0}\) identically, 
\(\mathop{\mathrm{div}}\bm{P}\) as given by 
\eqref{equ:FieTenStsDiv} then vanishes identically in this case. 
So for \(\msbi{S}\) left-minor skew-symmetric, i.e., \eqref{equ:SymSkwLef}, 
\(\bm{P}\) as defined by \eqref{equ:FieTenStsRie} is divergence-free. Without 
loss of computational generality, one can assume that \(\msbi{S}\) 
is both left- and right-minor skew-symmetric, i.e., 
\begin{equation}
\msbi{S}^{\mathrm{T}}\!\bm{A}
=\msbi{S}^{\mathrm{T}}\!\mathop{\mathrm{skw}}\bm{A}
\,,\quad
\msbi{S}\bm{A}
=\msbi{S}\mathop{\mathrm{skw}}\bm{A}
\,,
\end{equation}
via \eqref{equ:MajTraTenOrdFou}. 

Again analogous to \eqref{equ:StsPotSerForFluMea} and 
\eqref{equ:StsKirPioFirPotSerForCur}${}_{1,2}$, the above relations 
can be expressed in Fourier series form 
\begin{equation}
\begin{array}{rclcl}
\msbi{S}(\bm{x})
&=&
\skew3\bar{\msbi{S}}+\skew3\tilde{\msbi{S}}(\bm{x})
&=&
\skew3\hat{\msbi{S}}(\bm{0})
+\sum_{\bm{k}\neq\bm{0}}
e_{}^{\imath{}\bm{k}\cdot\bm{x}}
\skew3\hat{\msbi{S}}(\bm{k})
\,,\\
\mathop{\mathrm{div}}\msbi{S}(\bm{x})
&=&
\mathop{\mathrm{div}}\skew3\tilde{\msbi{S}}(\bm{x})
&=&
\sum_{\bm{k}\neq\bm{0}}
e_{}^{\imath{}\bm{k}\cdot\bm{x}}
\skew3\hat{\msbi{S}}(\bm{k})
\,\imath\bm{k}
\,.
\end{array}
\end{equation} 
From these follow in turn 
\begin{equation}
\begin{array}{rcl}
\bm{a}\cdot\bm{P}(\bm{x})\,\bm{b}
&=&
\sum_{\bm{k}\neq\bm{0}}
e_{}^{\imath{}\bm{k}\cdot\bm{x}}
\bm{a}\otimes\imath\bm{k}
\cdot
\skew3\hat{\msbi{S}}^{\mathrm{T}}(\bm{k})
\,\lbrack\bm{b}\otimes\imath\bm{k}\rbrack
\,,\\
\mathop{\mathrm{div}}\bm{P}(\bm{x})
&=&
\sum_{\bm{k}\neq\bm{0}}
e_{}^{\imath{}\bm{k}\cdot\bm{x}}
\skew3\hat{\msbi{S}}^{\mathrm{T}}(\bm{k})
\,\lbrack\mathop{\mathrm{skw}}\imath\bm{k}\otimes\imath\bm{k}\rbrack
\,\imath\bm{k}
\,,
\end{array}
\label{equ:DivStsKirPioFirFor}
\end{equation} 
via \eqref{equ:MajTraTenOrdFou}, \eqref{equ:FieTenStsRie}, and 
\eqref{equ:SymSkwLef}. Since 
\(\mathop{\mathrm{skw}}\imath\bm{k}\otimes\imath\bm{k}=\bm{0}\) 
identically, \(\mathop{\mathrm{div}}\bm{P}\) vanishes identically, and 
\(\bm{P}\) as determined by \eqref{equ:DivStsKirPioFirFor}${}_{1}$ 
is divergence-free. 

As stated above, in three dimensions, any tensor with left and right 
skew-symmetry such as \(\msbi{S}\) has 9 independent components 
\(S_{\!ijkl}=\bm{i}_{i}\wedge\bm{i}_{\!j}
\cdot
\msbi{S}\lbrack\bm{i}_{k}\wedge\bm{i}_{l}\rbrack\). This corresponds 
to the number of independent components of \(\bm{P}\) in three dimensions. 

\end{appendix}

\end{document}